\documentclass[14pt]{article}
\usepackage{amsmath,amsfonts,amssymb,theorem,verbatim,graphics,epsfig}

\newtheorem{lem}{Lemma}[section] 
\def\BEN{\begin{enumerate}}  \def\BI{\begin{itemize}}
\def\EEN{\end{enumerate}}   \def\EI{\end{itemize}}
   \def\sec{\section} 

\def\beq{\begin{eqnarray}} \def\eeq{\end{eqnarray}}

\def\eqn#1{\begin{equation}#1\end{equation}}

\def\al*#1{\begin{align*}#1\end{align*}}

\def\ga*#1{\begin{gather*}#1\end{gather*}}

\def\alat*#1#2{\begin{alignat*}{#1}#2\end{alignat*}}
\def\bea{\begin{eqnarray*}}
\def\eea{\end{eqnarray*}}
\def\ml*#1{\begin{multline*}#1\end{multline*}}

\def\P{{\mathbb P}}   \def\i{\infty}
\def\E{{\mathbb E}}   \def\R{{\mathbb R}}
\def\N{{\mathbb N}}

\def\te#1{\mathrm{e}^{#1}}

\def\I{\int}      
     \def\th{\theta}

     \def\ps{\psi}
 \def\q{\qquad} 
\def\F{\Phi}   
  \def\td{\text{\rm d}}
 
\newtheorem{Thm}{Theorem}

{\theorembodyfont{\normalfont}

}

\newcommand{\Exp}{{\rm I\hspace{-0.7mm}E}}

\newcommand{\proof}{{\it Proof\ }}

\newcommand{\exit}{{\mbox{\, \vspace{3mm}}}
\hfill\mbox{$\square$}}

\begin{document}
\title{De Finetti's
dividend problem and impulse control for a two-dimensional
 insurance risk process}

\author{Irmina Czarna%
\footnote{Department of Mathematics, University of Wroc\l
aw, pl. Grunwaldzki 2/4, 50-384 Wroc\l aw, Poland, e-mail:
czarna@math.uni.wroc.pl}\ \hspace{0.3cm} Zbigniew
Palmowski\footnote{Department of Mathematics, University of Wroc\l
aw, pl. Grunwaldzki 2/4, 50-384 Wroc\l aw, Poland, e-mail:
zbigniew.palmowski@gmail.com}}
\maketitle

\begin{center}
\begin{quote}
\begin{small}
{\bf Abstract.} Consider two insurance companies (or two branches of the same
company) that receive premiums at different rates and then
split the amount they pay in fixed proportions for each claim
(for simplicity we assume that they are equal).
We model the occurrence of claims according
to a Poisson process. The ruin is achieved when the
corresponding two-dimensional risk process first leaves the
positive quadrant.
We will consider two scenarios of the controlled process: refraction and impulse control.
In the first case the dividends are payed out when
the two-dimensional risk process exits the fixed region.
In the second scenario, whenever the process hits the horizontal line,
it is reduced by paying dividends to some fixed
point in the positive quadrant where it waits for the next claim to arrive.
In both models we calculate the discounted cumulative dividend payments until the ruin.
This paper is the first attempt to understand the effect of dependencies of two portfolios on the joint optimal strategy
of paying dividends. For example in case of proportional reinsurance one can observe the interesting phenomenon that choice of the optimal barrier depends on the initial reserves. This is in contrast with the one-dimensional
Cram\'{e}r-Lundberg model where the optimal choice of the barrier is uniform for all initial reserves.
\end{small}
\end{quote}
\end{center}

\sec{Introduction}\label{sec:intro}
In collective risk
theory the reserves process $X$ of an insurance company is modeled
by:
\begin{equation}\label{mdim}
X(t) = u + c t - S(t),
\end{equation}
where $u>0$ denotes the initial reserve and
\begin{equation}\label{zlpoisson}
S(t)=\sum_{i=1}^{N_t} U_i\end{equation}
is a compound Poisson process. We assume that  $U_i$ ($i=1,2,...$) are i.i.d. distributed claims (with distribution function $F$).
Throughout this paper we will assume that claims have an absolutely continuous distribution with density $f$.
The arrival process is a homogeneous Poisson process $N_t$ with intensity $\lambda$.
The premium income is modeled by a constant premium density $c$ and the net profit condition is then $c>\frac{\lambda}{\alpha}$,
where $\frac{1}{\alpha}=\Exp[U_1]<\infty$.\\
Recently, several authors have
studied extensions of classical risk theory towards a
multidimensional reserves model (\ref{mdim}),
where $X(t), x, c$ and $S(t)$ are vectors, with possible
dependence between the components of $S(t)$. Indeed, the
assumption of independence of risks may easily fail, for example
in the case of reinsurance, when incoming claims have an impact on
both companies at the same time. In general, one can also
consider situations where each claim event might induce more than
one type of claim in an umbrella policy (see Sundt \cite{Su}). For
some recent papers considering dependent risks, see Dhaene and
Goovaerts \cite{DG, DG2}, Goovaerts and Dhaene \cite{DG3}, M\"{u}ller
\cite{Mulla, Mullb}, Denuit et al. \cite{DGM}, Ambagaspitiya \cite{Am}, Dhaene and
Denuit \cite{DD}, Hu and Wu \cite{HW} and Chan et al. \cite{chan}.

In this paper we consider a particular
two-dimensional risk model in which two companies
receive premiums at rates $c_1$ and $c_2$ and then
split the
amount they pay in fixed proportions for each claim
(for simplicity we assume that they are equal).
That is,
\begin{equation}\label{mdim2}\underline{X}(t)=(X_1(t),X_2(t))=\left(u_1+c_1t-\sum_{i=1}^{N_t} U_i,u_2+c_2t-\sum_{i=1}^{N_t} U_i\right).
\end{equation}
Note that (\ref{mdim2}) models proportional reinsurance dependence.
The same two-dimensional risk process was already considered in Avram et al. \cite{APPinsmon, APPins, APPinsime}.
In Avram et al. \cite{APPins} under a
Cram\'er light-tailed  assumption for the claim size distribution,
the asymptotics of the ruin probability is derived when the
initial reserves of both companies tend to infinity.
In Avram et al. \cite{APPinsmon, APPinsime}, for the simplest particular
case of exponentially distributed claims, the Laplace transforms of few perpetual ruin
probabilities are identified. This allows to derive explicit expressions for these ruin probabilities.
These papers have not taken into account dividend payments.
The model (\ref{mdim2}) is also related with the joint steady-state workload in a two-node tandem with L\'{e}vy input
(see \cite{KellaWhitt, MaLi} and references therein).
The overview of two-dimensional models is given in \cite{Asbookruinprob}.

Without loss of generality, we can assume that the second company, to be called reinsurer, receives less
premium per amount paid out, that is,
\begin{equation}\label{c12}c_1>c_2.\end{equation}

We assume additionally that any premium income per unit time $c_i$ is larger than
the average amount claimed $\lambda \Exp[U_1]$. Thus, the two-dimensional surplus in
model (\ref{mdim2}) has
the unrealistic property that it converges to infinity on each co-ordinate
with probability one. In answer to this objection De Finetti
\cite{Fin} introduced the dividend barrier model for the one-dimensional model (\ref{mdim}), in which all
surpluses above a given level are transferred to a beneficiary.
Further, usually the payment of the dividends should be made in such a way as to optimize the expected
net present value of the total income of the shareholders from time zero until ruin.
Associated to each dividend payment could be a fixed
cost of size $K>0$. Then it is no longer feasible to pay out dividends at a certain rate. Therefore
the impulse control is considered instead where a dividend
strategy is given by the pairs of stopping times $(T_i,J_i)$. Variables $T_i$ ($i=1,2,\ldots$)
represent the times at which a dividend
payment is made and $J_i$ ($i=1,2,\ldots$) are positive random variables representing the sizes of the
dividend payments.

In
the mathematical finance and actuarial literature
there is a good
deal of work on dividend barrier models and the problem of finding
an optimal policy for paying out dividends. Gerber and Shiu
\cite{GerberEllias}, Grandits et al. \cite{GHS} and Jeanblanc and Shiryaev \cite{JeanShir}
consider the optimal dividend problem in a Brownian setting.
Irb\"{a}ck \cite{Irback}, Zhou \cite{Zhou}, Zajic \cite{Zajic}, Avram et al. \cite{APPdiv},
Kyprianou and Palmowski \cite{KyprPalm} and Loeffen \cite {Loeffen1} study
the constant barrier model for a classical and spectrally negative L\'{e}vy risk process. Azcue and
Muler \cite{AM} follow a viscosity approach to investigate optimal
reinsurance and dividend policies in the Cram\'{e}r-Lundberg
model. In fact the De Finetti's objective is unrealistic since in this model ruin
is sure and its "severity" ignored. Therefore several alternative objectives have been proposed
recently, involving a penalty at ruin, based on a function of the severity of ruin.
This kind of problems is considered in Loeffen and Renaud \cite{LRenaud}.

The impulse control literature is also vast.
An important type of strategy for the impulse control problems is a so-called $(a^-,a^+)$ policy
which is similar to the well known
$(s, S)$ policy appearing in inventory control models. The $(a^-,a^+)$ policy
is a strategy where at each time the reserves are above a certain level $a^+$,
a dividend payment is made which brings the reserves down to another level $a^-$.
No dividends are paid out when the reserves are below $a^-$.
A similar model of paying impulse dividends is considered in this paper where regulation of the risk process brings
it to some fixed point.
When the risk process is a Brownian motion plus drift, Jeanblanc and Shiryaev \cite{JeanShir} show that the
optimal strategy for the impulse control problem is the $(a^-,a^+)$ policy.
Paulsen \cite{Paulsen} considers the case of a risk process modeled by a diffusion process and
shows that under certain conditions the $(a^-,a^+)$ policy is optimal. Note that in
Paulsen \cite{Paulsen} this type of strategy is referred to a lump sum dividend barrier
strategy.
A similar model is considered in
Alvarez \cite{Alvarez}, Cadenillas et al. \cite{Choulli}.
Loeffen \cite{Loeffen2} models the reserves by a spectrally negative L\'{e}vy process and finds
conditions under which the $(a^-,a^+)$ strategy is optimal.
Avram et al. \cite{APPimpulse} find an optimal impulse strategy for a L\'{e}vy risk process with a polynomial penalty function.

For the thorough overview of one-dimensional dividend barrier models see also \cite{Schmbook}.

We follow the same construction of controlled process like in the one-dimensional case. That is, we define controlled process by
\begin{equation}\label{refracted}
 \underline{Y}(t)=(Y_1(t),Y_2(t))=\underline{X}(t)-\underline{L}(t),
\end{equation}
where $\underline{L}(t)=(L_1(t), L_2(t))$ and $L_i(t)$ ($i=1,2$) are
nondecreasing $\mathcal{F}_t$-adaptable processes for a natural filtration $\{\mathcal{F}_t\}_{\{t\geq 0\}}$ of $S(t)$ describing cumulative dividend payments up to time $t$
by $i$th company.

In this paper, analogously to the one-dimensional case, we consider two controlling mechanism: refraction and impulse payments.
In Section \ref{sec:first} we deal with the refracted process $\underline{Y}$ for which:
\begin{equation}\label{refracted2}
\underline{L}(t)=\left(\delta_1\int_0^t\mathbf{1}_{\{\underline{Y}(s) \in \mathcal{B}\}}\,ds, \; \delta_2\int_0^t\mathbf{1}_{\{\underline{Y}(s)\in \mathcal{B}\}}\,ds\right)
\end{equation}
describes the two-dimensional linear drift at rate
\begin{equation}\label{deltapositive}
\underline{\delta}=(\delta_1, \delta_2)>(0,0)\end{equation} which is subtracted
from the increments of the risk process whenever it enters the fixed set:
\begin{equation}\label{B}\mathcal{B}=\{(x,z): x,z\geq 0\quad\text{and}\quad z\geq  b-ax\}, \qquad a,b>0.\end{equation}
The case $\underline{\delta}=\underline{c}-\underline{a}$ for $\underline{c}=(c_1,c_2)$ and $\underline{a}=(-1,a)$ corresponds to the reflected risk process at the line $z=b-ax$.
Then the risk process starting from $\mathcal{B}^c \cup \mathcal{B}_0$, where
\begin{equation}\label{bar0}\mathcal{B}_0=\{(x,z): x,z\geq 0\quad\text{and}\quad z= b-ax\},\end{equation}
stays there up to the ruin time (see Figure 1).

Formally, the refracted process is defined as a solution of stochastic differential equation
$\underline{Y}(t)=\underline{X}(t) -\underline{\delta}\int_0^t \mathbf{1}_{\{\underline{Y}(t) \in \mathcal{B}\}}$
in the following way. Define stopping times $\vartheta_n$ and $S_n$ recursively as follows. We set $S_0 = 0$ and
for $n = 1, 2,\dots$
\begin{eqnarray*}\vartheta_n &= &\inf\left\{t >S_{n-1}: \underline{X}(t)-\underline{\delta}\sum_{i=1}^{n-1}(S_i-\vartheta_i)\in \mathcal{B}\right\}, \\
S_n &= & \inf\left\{t >\vartheta_n: \underline{X}(t)-\underline{\delta}\sum_{i=1}^{n-1}(S_i-\vartheta_i)-\underline{\delta}(t-\vartheta_n)\notin \mathcal{B}\right\}.
\end{eqnarray*}
Note that the difference between two consecutive times is strictly positive (except possibly for
$S_0$ and $\vartheta_1$). We now construct a solution $\underline{Y}(t)$  issued from $\underline{Y}(0)=\underline{u}$ by:
\begin{eqnarray*}
\underline{Y}(t)=\left\{\begin{array}{ll}
\underline{X}(t)-\underline{\delta}\sum_{i=1}^{n-1}(S_i-\vartheta_i)&\mbox{for $t\in [S_n, \vartheta_{n+1})$ and $n=0,1,\ldots$}\\
\underline{X}(t)-\underline{\delta}\sum_{i=1}^{n-1}(S_i-\vartheta_i)-\underline{\delta}(t-\vartheta_n)&\mbox{for $t\in [\vartheta_n, S_{n})$ and $n=1,2,\ldots$}.
\end{array}\right.
\end{eqnarray*}

We will also assume one technical assumption that the first insurance company (after refraction) will get eventually ruined, namely:
\begin{equation}\label{ruinaforsure}
c_1-\delta_1<0.
\end{equation}
Note that this approach introduces a dependence structure in the dividend payments. In particular, we allow some reserves to move from one company to another
in form of dividend payments (or to move reserves between two branches of one insurance company).
Besides, in the case of independent risks, the optimal strategy keeps process in the area determined by the optimal line barriers of each company.
In contrast, the strategy considered here is based on reflecting at the line.
It describes then the joint strategy between insurer and reinsurer taking into account the relationship between the individual strategies.

In Section \ref{sec:second} we consider impulse control, where the $L_i(t)$ ($i=1,2$)
are the cumulative payments made whenever the two-dimensional risk process hits the horizontal line $z=u_2$. The size of the $i$th payment
equals $J_i=X_1(T_i)-u_1+(c_1+c_2)e_\lambda^{(i)}$, where $T_i$ is the $i$th moment of hitting the line $z=u_2$ by the process $\underline{X}(t)$.
Here $e^{(i)}_\lambda$ is independent of $X$, and it is an exponential random variable
with intensity $\lambda$. Each impulse payment $X_1(T_i)-u_1$
reduces the reserves to some fixed levels $(u_1,u_2)$. Also the incoming premiums $(c_1+c_2)e_\lambda^{(i)}$ are paid out as dividends until the next claim arrives.
Moreover, by condition (\ref{c12}) each time the second company (or the branch of the company)
has at least $u_2$ reserves, the first company's reserves are greater than $u_1$.
Therefore the dividends payments are always made by the first company which has greater premium rate (by reducing the reserves to
level $u_1$).
Associated to each impulse dividend payment is a fixed
cost of size $K>0$. In this model the reserves of the second
branch of the insurance company serves as a random control mechanism.
It produces information
when dividends should be paid from the reserves of the first branch.
This is a very convenient policy of paying dividend for insurance companies that have branches related via proportional reinsurance.

In this paper we focus on finding the $n$th moment
\begin{equation}\label{Vn}
V_n(u_1,u_2)=V_n(\underline{u})= \Exp[D^n|\underline{X}(0)=\underline{u}]\end{equation}
of the discount cumulative dividend payments
\begin{equation}\label{D}
D=(1,1)\cdot \int_0^{\sigma} e^{-q t} \, d\underline{L}(t),
\end{equation}
made until the ruin time $\sigma=\inf{\{t>0:\min{(Y_1(t),Y_2(t))}< 0\}}$ where $q$ is the discount factor.
Above $(\alpha_1,\alpha_2)\cdot (\alpha_3,\alpha_4)=\alpha_1\alpha_3+\alpha_2\alpha_4$ denotes the scalar product of two vectors in the plane
and $\underline{u}=(u_1,u_2)$.
Note that $\sigma$ is the first time when the controlled risk process exits the positive quadrant and it corresponds to the first time when at least one company gets ruined.

The paper is organized as follows. In Section \ref{sec:first} we derive a  partial integro-differential equation
for $V_n$. In Section \ref{sec:firstb} we provide an explicit unique solution
in terms of an infinite series in the case of exponentially distributed claim sizes $U_i$.
Finally, in Section \ref{sec:second} we solve the second dividend problem with the impulse control.
Section \ref{concl} presents our conclusions.

\sec{Barrier control}\label{sec:first}

In this section, we consider for the risk process (\ref{mdim2})
a refracted process (\ref{refracted})-(\ref{refracted2}) and use the barrier set (\ref{B}).
We focus here on deriving a partial integro-differential equation for $V_n$ defined in (\ref{Vn})-(\ref{D}).

Denote $\delta_0=\delta_1+\delta_2$ and
$$\frac{\partial f}{\partial \underline{u}}(\underline{u})=\left(\frac{\partial g}{\partial u_1}(\underline{u}), \frac{\partial g}{\partial u_2}(\underline{u})\right)$$
for a general function $g$.

\begin{Thm}\label{Vneq}
Assume that density $f(x)$ of claim sizes is bounded. Then the partial derivatives $\frac{\partial V_n}{\partial u_i}(\underline{u})$ ($i=1,2$) exist and for $\underline{u}\in \mathcal{B}^{\rm c}$ the function
$V_n$ is the unique solution of the equation:
\begin{eqnarray}\label{1.1}
\underline{c}\cdot\frac{\partial V_n}{\partial \underline{u}}(\underline{u})-(\lambda +nq) V_n(\underline{u})
+\lambda \int_0^{\min(u_1,u_2)} V_n(\underline{u}-(1,1)v)f(v)  \,dv =0
\end{eqnarray}
with the boundary conditions:
\begin{equation}\label{boundarythm} n\delta_0\;V_{n-1}(\underline{u})=\left.\underline{\delta}\cdot \frac{\partial V_n}{\partial \underline{u}}\right|_{\underline{u}\in\mathcal{B}_0},\qquad \underline{u}\in\mathcal{B}_0,\end{equation}
\begin{equation}\label{uciekabariera}\lim_{b\to\infty}V_n(\underline{u})=0,\qquad \underline{u}\in\mathcal{B}^{\rm c},
\end{equation}
\begin{equation}\label{gornyrog}
V_n(0,b)=0.
\end{equation}
\end{Thm}
\begin{proof}
We define the proof into three steps: {\bf 1)} proving that the partial derivatives of $V_{n}(\underline{u})$ are well-defined; {\bf 2)} identifying the equation for
$V_{n}(\underline{u})$ and {\bf 3)} proving uniqueness of the solution of this equation.

{\bf Step 1.}
For $\underline{u}=(u_1,u_2)\in\mathcal{B}^{\rm c}$ we denote by
$F_m^{\underline{u}}$ the event under which exactly $m$ claims arrives until first hitting the line $\mathcal{B}_0$.
Let $\underline{X}(0)=\underline{u}$
and $i$th interarrival time
and $i$th claim equal $x_i$ and $z_i$ ($i=1,2,\ldots,m$), respectively.
Then for $d\underline{z}=dz_1\,dz_2\ldots dz_m$ and $d\underline{x}=dx_1\,dx_2\ldots dx_m$ we have:
\begin{eqnarray*}
&&V_n(\underline{u})=\sum_{m=0}^\infty
\int_{F^{\underline{u}}_{m}}\exp\{-qn(T_{\underline{u}, \underline{z},\underline{x}})\}g_n (u_1-(z_1+\ldots+z_m)+c_1 T_{\underline{u}, \underline{z},\underline{x}} )\\&&
\qquad\qquad\qquad\qquad 
\lambda^m e^{-\lambda x_1}\ldots e^{-\lambda x_m} f(z_1)\ldots f(z_m)d\underline{z}\,d\underline{x},
\end{eqnarray*}
where
$$g_n(u)=V_n(u, b-au),\qquad u\in[0,b/a]$$
is the $n$th moment of discounted cumulative dividends until ruin when the risk process starts at the barrier $\mathcal{B}_0$ and
$$T_{\underline{u}, \underline{z},\underline{x}}=\frac{b-au_1-u_2+(a+1)(z_1+\ldots+z_m)}{c_2+ac_1}$$ is the first time of hitting $\mathcal{B}_0$.
Note now that to prove that partial derivatives of $V_n$ are well-defined and series
$\sum_{n=1}^\infty \frac{(-y)^n}{n!}\frac{\partial}{\partial u_i}V_n(\underline{u})$ ($i=1,2$) are uniformly bounded
it suffices to prove that {\bf 1a)} $g$ is differentiable and that {\bf 1b)}$\sum_{n=1}^\infty \frac{(-y)^n}{n!}g_n^\prime(u)$ is uniformly bounded.

{\bf Step 1a.}
Denote by $C_m^{u}$ the event under which exactly $m$ claims arrives until the ruin
when $\underline{X}(0)=(u,b-au)$ ($u\in [0,b/a]$).
Let $D_{u, \underline{z},\underline{x}}$ is a deterministic value of cumulative discounted dividends conditionally on the event that $i$th inter arrival time
and $i$th claim equal $x_i$ and $z_i$ ($i=1,2,\ldots,m$), respectively.
Note that for $h>0$ we have:
\begin{eqnarray*}
\lefteqn{g_n(u+h)-g_n(u)}\\
&&=\sum_{m=1}^\infty\bigg(\int_{C^{u+h}_m}D^n_{u+h, \underline{z},\underline{x}} \lambda^m e^{-\lambda x_1}\ldots e^{-\lambda x_m} f(z_1)\ldots f(z_m)d\underline{z}\,d\underline{x}\\&&\quad-
\int_{C^{u}_m}D^n_{u, \underline{z},\underline{x}} \lambda^m e^{-\lambda x_1}\ldots e^{-\lambda x_m} f(z_1)\ldots f(z_m)d\underline{z}\,d\underline{x}\bigg)\\
&&=\sum_{m=1}^\infty\bigg(\int_{C^{u+h}_m\cap C^{u}_m}D^n_{u+h,\underline{z},\underline{x}} \lambda^m e^{-\lambda x_1}\ldots e^{-\lambda x_m} f(z_1)\ldots f(z_m)d\underline{z}\,d\underline{x}\\&&\quad-
\int_{C_m^{u+h}\cap C^{u}_m}D^n_{u, \underline{z},\underline{x}} \lambda^m e^{-\lambda x_1}\ldots e^{-\lambda x_m} f(z_1)\ldots f(z_m)d\underline{z}\,d\underline{x}\bigg)\\&&\quad+
\sum_{m=1}^\infty\bigg(\int_{C^{u+h}_m\setminus C^{u}_m}D^n_{u+h, \underline{z},\underline{x}} \lambda^m e^{-\lambda x_1}\ldots e^{-\lambda x_m} f(z_1)\ldots f(z_m)d\underline{z}\,d\underline{x}\\&&\quad-
\int_{C^{u}_m\setminus C^{u+h}_m}D^n_{u, \underline{z},\underline{x}} \lambda^m e^{-\lambda x_1}\ldots e^{-\lambda x_m} f(z_1)\ldots f(z_m)d\underline{z}\,d\underline{x}\bigg).
\end{eqnarray*}
The first series above equals $0$. It follows from observing parallel paths of the regulated process $\underline{Y}$
starting at $(u+h,b-a(u+h))$ and $(u, b-au)$ until ruin which is the same for both trajectories. Hence $D_{u, \underline{z},\underline{x}}=D_{u+h, \underline{z},\underline{x}}$. The only positive difference appears
when both trajectories have different moments of ruin. That is,
\begin{eqnarray}
\lefteqn{g_n(u+h)-g_n(u)}\nonumber\\&&=
\sum_{m=1}^\infty\bigg(\int_{C^{u+h}_m\setminus C^{u}_m}D^n_{u+h, \underline{z},\underline{x}} \lambda^m e^{-\lambda x_1}\ldots e^{-\lambda x_m} f(z_1)\ldots f(z_m)d\underline{z}\,d\underline{x}
\nonumber\\&& \quad-
\int_{C^{u}_m\setminus C^{u+h}_m}D^n_{u, \underline{z},\underline{x}} \lambda^m e^{-\lambda x_1}\ldots e^{-\lambda x_m} f(z_1)\ldots f(z_m)d\underline{z}\,d\underline{x}\bigg)
\nonumber\\
&&=\sum_{m=1}^{\infty}\sum_{k=1}^{m-1}\int_{-h}^0\int_0^{b}\P(Y_1(\sigma)\in dx, Y_2(\sigma)\in dw, C^u_k|\underline{Y}(0)=(u, b-au))\nonumber
\\&&\qquad\cdot \E[D^n1_{(C^{u+h}_m\cap C^{u}_k)}|\underline{Y}(0)=(u+h, b-a(u+h))]\nonumber\\
&&+\sum_{m=1}^{\infty}\sum_{k=m+1}^{\infty}\int_{-ah}^0\int_0^{b/a}\P(Y_1(\sigma)\in dw, Y_2(\sigma)\in dx, C_k^u|\underline{Y}(0)=(u+h, b-a(u+h)))\nonumber
\\&&\qquad\cdot \E[D^n1_{(C^{u+h}_m\cap C^{u}_k)}|\underline{Y}(0)=(u+h, b-a(u+h))]\nonumber
\\&&-
\sum_{m=1}^{\infty}\sum_{k=1}^{m-1}\int_{-ah}^0\int_0^{b/a}\P(Y_1(\sigma)\in dw, Y_2(\sigma)\in dx, C^{u+h}_k|\underline{Y}(0)=(u+h, b-a(u+h)))\nonumber\\&&\qquad\cdot
\E[D^n1_{(C^{u}_m\cap C^{u+h}_k)}|\underline{Y}(0)=(u, b-au)]
\nonumber\\&&-
\sum_{m=1}^{\infty}\sum_{k=m+1}^{\infty}\int_{-h}^0\int_0^{b}\P(Y_1(\sigma)\in dx, Y_2(\sigma)\in dw, C^{u+h}_k|\underline{Y}(0)=(u+h, b-a(u+h)))\nonumber\\&&\qquad\cdot
\E[D^n1_{(C^{u}_m \cap C^{u+h}_k)}|\underline{Y}(0)=(u, b-au)],\label{suma}\end{eqnarray}
where the first integrals $\int_{-h}^0$ and $\int_{-ah}^0$ are taken with respect to $dx$.
Note that
under our assumptions $\underline{Y}(t)$ is a Feller process and that
\begin{eqnarray}
&&\max_{\underline{u}\in\mathcal{B}^{\rm c}\cup
\mathcal{B}_0}
V_{n}(\underline{u})= \max_{\underline{u}\in\mathcal{B}^{\rm c}\cup
\mathcal{B}_0}
\E\left[(1,1)\int_0^\infty e^{-qs}\,d\underline{L}_s\right]^n
= \max_{\underline{u}\in\mathcal{B}^{\rm c}\cup
\mathcal{B}_0}\E\left[(1,1)q\int_0^\infty \underline{L}_s e^{-qs}\,ds\right]^n\nonumber\\
&&\quad\quad
\leq  \max_{\underline{u}\in\mathcal{B}_0}\left[q\int_0^\infty (\delta_1+\delta_2)s e^{-qs}\,ds\right]^n\leq \frac{(\delta_1+\delta_2)}{q}^n.\label{boudnthmoment}
\end{eqnarray}
Furthermore,
\begin{eqnarray*}
&&\frac{1}{h}\int_{-h}^0\int_0^{b}\P(Y_1(\sigma)\in dx, Y_2(\sigma)\in dw, C_k^u|\underline{Y}(0)=(u, b-au))
\\&&=\frac{1}{h}\int_0^{b/a}\int_0^{b}\int_z^{z+h}\P(Y_1(\sigma-)\in dz, Y_2(\sigma-)\in dw+s, C_k^u|\underline{Y}(0)=(u, b-au))f(s)\,ds.
\end{eqnarray*}
The latter expression converges as $h\downarrow 0$. A similar result could be derived for the other terms appearing in (\ref{suma}).
Hence by the dominated convergence theorem and (\ref{boudnthmoment}) the derivative $g^\prime_n(u)$ exists for each $n=1,2,\ldots$ and $u\in (0,b/a]$.

{\bf Step 1b.}
Moreover, $g_n^\prime(u)$ and then also $\frac{\partial}{\partial u_i}V_n(\underline{u})
$ ($i=1,2$) are continuous functions. Note also that:
$$g^\prime_n(u)\leq 4\max\{a,1\}\max_s f(s)\max_{\underline{u}\in\mathcal{B}^{\rm c}\cup\mathcal{B}_0}
V_{n}(\underline{u}).$$

By Fubini's theorem:
\begin{equation}\label{serdiff}
\sum_{n=1}^\infty \frac{(-y)^n}{n!}g_n^\prime(u)\leq 4\max\{a,1\}\max_s f(s)\max_{\underline{u}\in\mathcal{B}^{\rm c}\cup\mathcal{B}_0}M(\underline{u},y)\leq 4\max\{a,1\}\max_s f(s),
\end{equation}
where for $y\geq 0$
$$M(\underline{u},y)=\Exp\left[e^{-yD}|\underline{X}(0)=\underline{u}\right]$$
is the Laplace transform of $D$ and
\begin{equation}\label{bb}
0\leq M(\underline{u},y)\leq 1.
\end{equation}

Since the series
$
\sum_{n=1}^\infty \frac{(-y)^n}{n!}\frac{\partial}{\partial u_i}V_n(\underline{u})
$ ($i=1,2$) are uniformly bounded, the partial derivatives of $M$ are well defined.

{\bf Step 2.}
By the Strong Markov Property we have that for $\underline{u}\in \mathcal{B}^c$:
\begin{eqnarray}
\nonumber M(\underline{u},y)&=&(1-\lambda dt)M(\underline{u}+\underline{c}dt,ye^{-q dt})\\
\nonumber &&+\lambda dt \int_0^{\min(u_1+c_1 dt,u_2+c_2 dt)} M(\underline{u}+\underline{c}dt-(1,1)v,ye^{-q dt}) f(v) \, d v\\
&&+\lambda dt  \int_{\min(u_1+c_1 dt,u_2+c_2 dt)}^{\infty} f(v)\,dv +o(dt).\label{fala}
\end{eqnarray}

Then a Taylor expansion and collection of terms of order $dt$ yields
\begin{eqnarray}
&&\underline{c}\cdot \frac{\partial M}{\partial \underline{u}}(\underline{u},y)-\lambda M(\underline{u},y)
-q y \frac{\partial M}{\partial y}(\underline{u},y)
+\lambda \int_0^{\min(u_1,u_2)} M(\underline{u}-(1,1)v,y)  f(v)\, d v\nonumber\\&& \hspace{6cm}+\lambda (1-F(\min(u_1,u_2)))=0.\label{1.0}
\end{eqnarray}
Similar considerations will produce the following boundary equation for $\underline{u}\in\mathcal{B}_0$:
\begin{eqnarray}
\nonumber M(\underline{u},y)&=&(1-\lambda dt)e^{-y\delta_0 dt}M(\underline{u}+(\underline{c}-\underline{\delta})dt,ye^{-q dt})\\
\nonumber &&+\lambda dt \int_0^{\min(u_1+(c_1-\delta_1)dt, u_2 +(c_2- \delta_2)dt)}e^{-y\delta_0dt} \\
&&\hspace{2cm}M(\underline{u}-(\underline{c}-\underline{\delta})dt-(1,1)v,ye^{-q dt})f(v)  \, d v\nonumber\\
\nonumber &&+\lambda dt \textrm{ } e^{-y\delta_0dt} \int_{\min(u_1+(c_1- \delta_1)dt, u_2+(c_2- \delta_2)dt)}^{\infty}f(v) \, d v +o(dt)
\end{eqnarray}
which implies:
\begin{eqnarray}\label{talaga}
\nonumber \left.(\underline{c}-\underline{\delta})\cdot\frac{\partial M}{\partial \underline{u}}(\underline{u},y)\right|_{\underline{u}\in\mathcal{B}_0}-(y\delta_0+\lambda) M(\underline{u},y)
-\left.q y \frac{\partial M}{\partial y}(\underline{u},y)\right|_{\underline{u}\in\mathcal{B}_0}\\
+\lambda \int_0^{\min(u_1, u_2)} M(\underline{u}-(1,1)v,y)  f(v)\, d v +\lambda (1-F(\min(u_1, u_2)))=0.
\end{eqnarray}
Observe now that, by the lack of memory of exponential distribution between arrivals of claims, for $\underline{u}\in\mathcal{B}_0$ and $h>0$ we have,
$$e^{-(\lambda+q) h}M(\underline{u}, y)=M(\underline{u}-\underline{c}h, y)-\int_0^h\lambda e^{-(\lambda+q) t}\int_0^\infty M(\underline{u}-\underline{c}h+\underline{c}t-z, y)f(z)\,dt\,dz.$$
Taking the operator $\underline{c}\frac{\partial}{\partial \underline{u}}$ on both sides of the above equation, applying in the next step equation (\ref{1.0}) for $M(\underline{u}-\underline{c}h, y)$
and then taking limit $h\downarrow 0$ will prove that equation (\ref{1.0}) also holds for $\underline{u}\in\mathcal{B}_0$.
Thus from (\ref{talaga}) we finally derive the following boundary condition for $\underline{u}\in\mathcal{B}_0$:
\begin{equation}\label{talaga2} -y\delta_0M(\underline{u},y)=\left.\underline{\delta}\frac{\partial M}{\partial \underline{u}}(\underline{u},y)\right|_{\underline{u}\in\mathcal{B}_0}.\end{equation}

Using now representation $M(\underline{u},y)=1+\sum_{n=1}^{\infty} \frac{(-y)^n}{n!} V_n(\underline{u})$
and equating the coefficient of $(-y)^n$ in (\ref{1.0}) completes the proof of (\ref{1.1})-(\ref{boundarythm}).
Moreover, conditions (\ref{uciekabariera})-(\ref{gornyrog}) follow straightforwardly from the definition of $D$ in (\ref{D}) and
inequality (\ref{ruinaforsure}).

{\bf Step 3.}
We will prove the uniqueness of the solution using arguments similar the ones in Gerber \cite{Gerber81}.
It suffices to prove the uniqueness of the solution of (\ref{1.0}) and (\ref{talaga2}).
For that purpose, we define
the operator $\mathcal{A}$ by
\begin{eqnarray*}
\lefteqn{\mathcal{A}g (\underline{u},y):=\lambda\int_0^T e^{-\lambda t}\int_0^{\min(u_1+c_1 t,u_2+c_2 t)} g(\underline{u}+\underline{c}t-(1,1)v,ye^{-qt}) f(v) \, dv\,dt
}\\&&+\lambda \Exp\int_T^\sigma e^{-\lambda t}\exp\left\{-y\delta_0\int_T^t e^{-qs}\,ds\right\}\int_0^{\min(u_1+(c_1-\delta_1) (t-T),u_2+(c_2-\delta_2)(t-T))}\\&&
\hspace{4cm} g(\underline{u}+(\underline{c}-\underline{\delta})(t-T)-(1,1)v,ye^{-qt}) f(v) \, dv
\,dt,
\end{eqnarray*}
where $T=(b-au_1-u_2)/(c_2+ac_1)$ for $\underline{u}\in \mathcal{B}^c$ is the first time of getting to the linear barrier $\mathcal{B}_0$ defined in (\ref{bar0}).
Note that in $\mathcal{A}$ each increment can be interpreted as a conditioning on
whether a claim occurs before the surplus process hits the barrier ($t <T$) or after
this event (in which case we have an additional term $\exp\left\{-y\delta_0\int_T^t e^{-qs}\,ds\right\}$ representing the discounted dividends
paid until the claim occurs). The solution $M$ of (\ref{1.0}) and (\ref{talaga2}) is
a fixed point of the integral operator $\mathcal{A}$. For two functions $g_1$ and $g_2$
we have:
\begin{eqnarray*}
\lefteqn{|\mathcal{A}g_1(\underline{u},y)-\mathcal{A}g_2(\underline{u},y)|}
\\&& \leq ||g_1(\underline{u},y)-g_2(\underline{u},y)||_\infty\left(\lambda\int_0^Te^{-\lambda t}\,dt\right.\\&&\left.\hspace{4cm}+
\lambda\int_T^\infty e^{-\lambda t}\exp\left\{-y\delta_0\int_T^t e^{-qs}\,ds\right\}\,dt\right)\\&&<||g_1(\underline{u},y)-g_2(\underline{u},y)||_\infty,
\end{eqnarray*}
where $||\cdot||_\infty$ is the supremum norm over $\underline{u}\in \R^2$ and $y\in \R_+$.

Thus it follows that $\mathcal{A}$ is a contraction and by
Banach's fixed point theorem and (\ref{bb}) the solution of (\ref{1.1})-(\ref{gornyrog}) is unique.
\exit
\end{proof}


\sec{An explicit solution for the exponential claims size and reflection}\label{sec:firstb}

In this section we will find the unique solution of (\ref{1.1})-(\ref{gornyrog}) for $n=1$, exponential claim size $F(v)=1-e^{-\alpha v}$ and reflection at line
$z=b-ax$, that is $\underline{\delta}=\underline{c}-\underline{a}$, where $\underline{a}=(-1,a) $ (see Figure 1).
We do not manage to find any other solutions of (\ref{1.1})-(\ref{gornyrog}).

Note that by (\ref{deltapositive}) we have
\begin{equation}\label{ac2}
c_2>a.
\end{equation}
Without loss of generality we will assume that $u_1<u_2$. In the case $u_1\geq u_2$ the solution of (\ref{1.1})-(\ref{boundarythm})
could be modified in the obvious way.

\begin{figure}[t] \label{piece}
\centering
\resizebox{.40 \textwidth}{!}{

\centerline{\setlength{\unitlength}{
3947sp}
\begingroup\makeatletter\ifx\SetFigFont\undefined%
\gdef\SetFigFont#1#2#3#4#5{%
  \reset@font\fontsize{#1}{#2pt}%
  \fontfamily{#3}\fontseries{#4}\fontshape{#5}%
  \selectfont}%
\fi\endgroup%
\begin{picture}
(5000, 4000)(2000,-7502)

\thinlines
{\put(2101,464){\line( 0,-1){7875}}
}%
{\put(2101,-7411){\line( 1, 0){5000}}
}%
{\put(
7101
,-7411){\vector( 1, 0){675}}
}%
{\put(2101,464){\vector( 0, 1){300}}
}%
{\put(2101,239){\line( 0,-1){ 75}}
}%

{\put(2851,-5836){\line( 1, 0){300}}
\put(3151,-5836){\line(-1, 0){ 75}}
}%
{\put(3301,-5836){\line( 1, 0){300}}
}%
{\put(3601,-5836){\line( 0,-1){225}}
}%
{\put(3601,-6211){\line( 0,-1){225}}
}%
{\put(3601,-6661){\line( 0,-1){300}}
}%
{\put(3601,-7186){\line( 0,-1){225}}
}%

{\put(5101,-5311){\line(-3,-4){225}}
}%
{\put(4777,-5743){\line(-3,-4){225}}
}%
{\put(4477,-6193){\line(-3,-4){225}}
}%

{\put(2827,-3193){\line(-3,-4){225}}
}%
{\put(3151,-2761){\line(-3,-4){225}}
}%
{\put(3451,-2311){\line(-3,-4){225}}
}%
{\put(3751,-1861){\line(-3,-4){225}}
}%
\thicklines
{\put(6336,-5738){\vector(-2, 3){2584.615}}
}%
\thinlines

{\put(
6340,-5730){\line( 3,-4){
1260}}
}%

{\put(3751,-1861){\line(-2, 3){1615.385}}
}%
\thicklines
{\put(3601,-5836){\line( 3, 1){1507.500}}
}%
\thinlines
{\put(4102,-2968){\line(-3,-4){225}}
}%
{\put(3802,-3343){\line(-3,-4){225}}
}%
{\put(3541,-3691){\line(-3,-4){225}}
}%
{\put(3229,-4132){\line(-3,-4){225}}
}%
{\put(2941,-4516){\line(-3,-4){225}}
}%
{\put(2629,-4882){\line(-3,-4){225}}
}%
{\put(2329,-5257){\line(-3,-4){225}}
}%
\thicklines
{\put(2622,-3497){\line( 3, 1){1507.500}}
}%
{\put(4276,-6511){\line( 5, 2){2004.310}}
}%
\thinlines
{\put(2101,-5836){\line( 1, 0){225}}
}%
{\put(2476,-5836){\line( 1, 0){225}}
}%
\put(3451,-7711){\makebox(0,0)[lb]{\smash{{\SetFigFont{12}{14.4}{\rmdefault}{\mddefault}{\updefault}{$u_1$}%
}}}}
\put(1801,-5911){\makebox(0,0)[lb]{\smash{{\SetFigFont{12}{14.4}{\rmdefault}{\mddefault}{\updefault}{$u_2$}%
}}}}
\put(5101,-3361){\makebox(0,0)[lb]{\smash{{\SetFigFont{12}{14.4}{\rmdefault}{\mddefault}{\updefault}{$\underline{a}=(-1,a)$}%
}}}}
\put(5251,-6361){\makebox(0,0)[lb]{\smash{{\SetFigFont{12}{14.4}{\rmdefault}{\mddefault}{\updefault}{$\underline{Y}(t)$}%
}}}}
\end{picture}%
}
}

\caption{Controlled two-dimensional risk process. }
\end{figure}
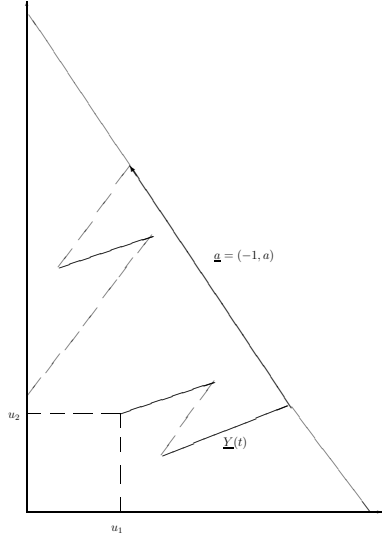

We now introduce some notations and gather some necessary prerequisites for the main result.

For $m\in \R$ let
\begin{equation}\label{gamma2zero} \gamma_{2,0}(m)=\frac{-[m(\alpha c_1 -q -\lambda)+\alpha c_2 -q -\lambda]+\sqrt{\Delta_{\gamma_{2,0}}(m)}}{2((m^2+m)c_1 +c_2+mc_2)}>0,\end{equation}
 where \begin{equation}\label{Deltagammadwa}\Delta_{\gamma_{2,0}}(m)=[m(\alpha c_1 -q -\lambda)+\alpha c_2 -q -\lambda]^2+4\alpha q[(m^2+m)c_1 +c_2+mc_2)]>0.\end{equation}
Moreover, for
\begin{equation}\label{aprime}a^\prime=\frac{a-c_2}{c_1+1}<0\end{equation}
denote
\begin{equation}\label{gamma2zero}\gamma_{2,0}=\gamma_{2,0}(a),\qquad \gamma_{2,0}^\prime=\gamma_{2,0}(a^\prime).\end{equation}

Each equation with respect to $\gamma$:
\begin{eqnarray}\label{sqeq}
c_1  \gamma^2 + c_2  \gamma_{2,k}^2 +(c_1+c_2) \gamma  \gamma_{2,k} + (\alpha c_1 -q -\lambda)\gamma+
(\alpha c_2 -q-\lambda)\gamma_{2,k} -\alpha q   =0,
\end{eqnarray}
\begin{eqnarray}\label{sqeqprime}
c_1  \gamma^{2} + c_2  \gamma_{2,k}^{2,\prime} +(c_1+c_2) \gamma  \gamma_{2,k}^\prime + (\alpha c_1 -q -\lambda)\gamma+
(\alpha c_2 -q-\lambda)\gamma_{2,k}^\prime -\alpha q   =0,
\end{eqnarray}
has two solutions $\gamma_{1,k}$, $\gamma_{3,k}<\gamma_{1,k}$ and $\gamma_{1,k}^\prime$, $\gamma_{3,k}^\prime<\gamma_{1,k}^\prime$ ($k=0,1,2,\ldots$),
respectively.
Note that   $\gamma_{2,0}$ and $\gamma_{2,0}^\prime$ were chosen in such a way that
\begin{equation}\label{Lemma31ia}
\gamma_{1,0}=a \gamma_{2,0}>0,\qquad \gamma_{1,0}^\prime=a^\prime\gamma_{2,0}^\prime>0.
\end{equation}
Moreover, for $k=0,1,2,\ldots$ we will consider additional two equations:
\begin{equation}\label{Lemma31i}
\gamma_{1,k+1}=a \gamma_{2,k+1}+\gamma_{3,k}-a \gamma_{2,k}
\end{equation}
and
\begin{equation}\label{Lemma31iprime}
\gamma_{1,k+1}^\prime=a \gamma_{2,k+1}^\prime +\gamma_{3,k}^\prime-a \gamma_{2,k}^\prime.
\end{equation}
Given $\gamma_{2,k}$, $\gamma_{3,k}$ and $\gamma_{2,k}^\prime$, $\gamma_{3,k}^\prime$,
we put expressions (\ref{Lemma31i}) and (\ref{Lemma31iprime}) for $\gamma_{1,k+1}$ and $\gamma_{1,k+1}^\prime$ into (\ref{sqeq}) and (\ref{sqeqprime}), respectively.
In this way we derive square equations for $\gamma_{2,k+1}$ and $\gamma_{2,k+1}^\prime$ (we choose
their biggest positive roots). Having $\gamma_{2,k+1}$, $\gamma_{2,k+1}^\prime$ we find $\gamma_{1,k+1}$, $\gamma_{3,k+1}$ and
$\gamma_{1,k+1}^\prime$, $\gamma_{3,k+1}^\prime$ as a solutions of (\ref{sqeq}) and (\ref{sqeqprime}), respectively.
In the next lemma we show that all these quantities are well defined and prove that $\gamma_{2,k}$, $\gamma_{2,k}^\prime$, $\gamma_{3,k}$, $\gamma_{3,k}^\prime$
create increasing sequences.

\begin{lem}\label{recc} Let $k\in\N\cup\{0\}$. Then
\begin{description}
\item{(i)} $\gamma_{2,0}>0$ and  $\gamma_{2,0}^\prime>0$;
\item{(ii)} $\gamma_{i,k}$ and $\gamma_{i,k}^\prime$ ($i=1,2,3$, $k=0,1,2,\dots$) satisfying  (\ref{sqeq}), (\ref{Lemma31i})  and (\ref{sqeqprime}), (\ref{Lemma31iprime}), respectively, always exist;
\item{(iii)} $\gamma_{2,k+1}>\gamma_{2,k}>0 $, $\gamma_{1,k}>\gamma_{3,k}$ and $\gamma_{3,k+1}<\gamma_{3,k}<0$; similarly $\gamma_{2,k+1}^\prime>\gamma_{2,k}^\prime>0 $, $\gamma_{1,k}^\prime>\gamma_{3,k}^\prime$ and $\gamma_{3,k+1}^\prime<\gamma_{3,k}^\prime<0$.
\end{description}
\end{lem}
\proof
Putting $\gamma_{1,0}=a\gamma_{2,0}$  into (\ref{sqeq}) gives the equation:
$$\mathcal{I}_1:=((a^2+a)c_1 +(1+a)c_2)\gamma_{2,0}^2+(a(\alpha c_1 -q -\lambda)+\alpha c_2 -q -\lambda)\gamma_{2,0}-\alpha q =0$$
which has one positive solution $\gamma_{2,0}$ given in (\ref{gamma2zero}) and one negative solution, since
$\frac{-\alpha q}{(a^2+a)c_1 +(1+a)c_2}<0$.
Moreover,  by (\ref{c12}) we have \begin{equation}\label{wspaprim}(a^{2,\prime}+a^\prime)c_1 +(1+a^\prime)c_2>0\end{equation} and hence
equation
$$((a^{2, \prime}+a^\prime)c_1 +(1+a^\prime)c_2)\gamma_{2,0}^{2,\prime}+(a^\prime(\alpha c_1 -q -\lambda)+\alpha c_2 -q -\lambda)\gamma_{2,0}^\prime-\alpha q =0$$
has also one positive solution $\gamma_{2,0}^\prime$.
This completes the proof of (i).
Note that $\gamma_{3,0}$ and $\gamma_{3,0}^\prime$ are also well-defined since for (\ref{sqeq}):
\begin{eqnarray}\nonumber
\Delta_{\gamma_{1,0}}:=
(c_1-c_2)^2\gamma_{2,0}^2+2\gamma_{2,0}(c_1-c_2)(\alpha c_1+\lambda+q)+(\alpha c_1-\lambda-q)^2+4c_1\alpha q>0
\end{eqnarray}
and
\begin{eqnarray}\nonumber
(c_1-c_2)^2\gamma_{2,0}^{2,\prime}+2\gamma_{2,0}^\prime(c_1-c_2)(\alpha c_1+\lambda+q)+(\alpha c_1-\lambda-q)^2+4c_1\alpha q>0.
\end{eqnarray}
From (\ref{sqeq}) we have that
\begin{equation}\label{neg}\mathcal{I}_2:=c_1\gamma_{1,0} \gamma_{3,0}=c_2  \gamma_{2,0}^2+(\alpha c_2-q-\lambda) \gamma_{2,0}-\alpha q<0, \end{equation}
where the last inequality follows from the form of $\mathcal{I}_1$. Indeed, then
$$
\mathcal{I}_2=-\gamma_{2,0}[((a^2+a)c_1 +ac_2)\gamma_{2,0}+a(\alpha c_1 -q -\lambda)]$$
is negative since by (i) $\gamma_{2,0}>0$ and by (\ref{gamma2zero}) and (\ref{gamma2zero}):
\begin{eqnarray}
\nonumber \lefteqn{\gamma_{2,0}-\frac{-a(\alpha c_1 -q -\lambda)}{(a^2+a)c_1 +ac_2}}\\&&=
 \nonumber (c_1-c_2)( a\alpha c_2+\lambda a^2+\lambda a+q a+q a^2)+a(a^2+a)c_1(\alpha c_1-\lambda)\\
&&\quad -qa(a^2+a)c_1 + \sqrt{\Delta_{\gamma_{2,0}}}((a^2+a)c_1 +ac_2)>0.\label{dodrazjeszcze}
\end{eqnarray}
The inequality (\ref{dodrazjeszcze}) is a consequence of the assumption (\ref{c12}),
the net profit condition $c_1>\frac{\lambda}{\alpha}$ and the inequality
$$ \sqrt{\Delta_{\gamma_{2,0}}} > qa$$
which is by (\ref{Deltagammadwa}) equivalent to
\begin{eqnarray*}
\lefteqn{\Delta_{\gamma_{2,0}}-q^2a^2=(a\alpha c_1-\lambda a +\alpha c_2 -q-\lambda)^2}\\&&+
2qa(a\alpha c_1+\alpha c_2+\lambda a+\lambda +q+2\alpha c_1)+4\alpha q c_2 >0.
\end{eqnarray*}
Then by (\ref{Lemma31ia}) and (\ref{neg}),
\begin{equation}\label{czteryzero} \gamma_{3,0}<0.\end{equation}
By (\ref{ac2}) we have $a-c_2\leq 0$ and hence by (\ref{aprime}) and (\ref{Lemma31ia}) $\gamma_{3,0}^\prime<\gamma_{1,0}^\prime\leq0 $.

Assume now that for some $k\in \N\cup\{0\}$ the quantities $\gamma_{2,k}>0, \gamma_{3,k}<0$ exist.
The solutions of equation
(\ref{sqeq}) then equals
\begin{eqnarray}
\gamma_{3,k}&=&\frac{-(c_1+c_2)\gamma_{2,k}+\lambda+q-\alpha c_1-\sqrt{\Delta_{\gamma_{1,k}}}}{2c_1}\label{gama3}
\\ \gamma_{1,k}&=&\frac{-(c_1+c_2)\gamma_{2,k}+\lambda+q-\alpha c_1+\sqrt{\Delta_{\gamma_{1,k}}}}{2c_1},\label{gama1}
\end{eqnarray}
where
\begin{eqnarray*}
\Delta_{\gamma_{1,k}}
= (c_1-c_2)^2\gamma_{2,k}^2+2\gamma_{2,k}(c_1-c_2)(\alpha c_1+\lambda+q)+(\alpha c_1-\lambda-q)^2+4c_1\alpha q.
\end{eqnarray*}
Note that $\gamma_{1,k}>\gamma_{3,k}$ and that $\Delta_{\gamma_{1,k}}>0 $.
Moreover, putting (\ref{Lemma31i}) into (\ref{sqeq})
produces the equation for $\gamma_{2,k+1}$:
\begin{eqnarray}&&((a^2+a)c_1 +(1+a)c_2)\gamma_{2,k+1}^2\nonumber\\&&+[(\gamma_{3,k}-a\gamma_{2,k})(2c_1a+c_1+c_2)-(\lambda +q)(1+a)+\alpha(ac_1+c_2)]\gamma_{2,k+1}\nonumber\\&&
+(c_1(\gamma_{3,k}-a\gamma_{2,k})^2+(c_1\alpha-\lambda -q)(\gamma_{3,k}-a\gamma_{2,k})-\alpha q)=0\label{gamma2k}
\end{eqnarray}
which has a solution since:
\begin{eqnarray}
\lefteqn{\Delta_{\gamma_{2,k+1}}= [\alpha(ac_1+c_2)-(\lambda +q)(1+a)]^2}\nonumber\\&&+4\alpha q((a^2+a)c_1 +(1+a)c_2)
+(\gamma_{3,k}-a\gamma_{2,k})^2(c_1-c_2)^2\nonumber\\&&+2(\gamma_{3,k}-a\gamma_{2,k})(c_2-c_1)((\lambda+q)(1+a)
+\alpha(c_2+ac_1))>0.\label{deltagamma2k1}
\end{eqnarray}
Furthermore,
\begin{eqnarray}
\lefteqn{\gamma_{2,k+1}-\gamma_{2,k}
= \sqrt{\Delta_{\gamma_{2,k+1}}} + (2c_1a+c_1+c_2 )\sqrt{\Delta_{\gamma_{1,k}}}}\nonumber\\&&
+\frac{(c_1-c_2)^2}{2c_1}\gamma_{2,k} + \frac{(c_1-c_2)}{2}(\alpha +\lambda+q)>0.\label{gammapositive diff}
\end{eqnarray}
Hence $\gamma_{2,k+1}>0$ for $k=-,1,2,\ldots$ since $\gamma_{2,0}>0$ and
\begin{equation}\label{deltagamma1k}\Delta_{\gamma_{1,k+1}}=(c_1-c_2)^2\gamma_{2,k+1}^2+2\gamma_{2,k+1}(c_1-c_2)(\alpha c_1+\lambda+q)+(\alpha c_1-\lambda-q)^2+4c_1\alpha q>0\end{equation}
which means that there exist two solutions $\gamma_{1,k+1}, \gamma_{3,k+1}<\gamma_{1,k+1}$
of equation (\ref{sqeq}).
 Similarly, by (\ref{deltagamma2k1}) and (\ref{gammapositive diff}), using the same arguments we can prove that $\gamma_{2,k+1}^\prime>\gamma_{2,k}^\prime$ and that that there exist two solutions $\gamma_{1,k+1}^\prime, \gamma_{3,k+1}^\prime<\gamma_{1,k+1}^\prime$
of equation (\ref{sqeqprime}). This completes the proof of (ii).
To prove (iii) note first that:
\begin{eqnarray}\label{gamma3kmon}
\gamma_{3,k+1}-\gamma_{3,k}
=(c_1+c_2)(\gamma_{2,k}-\gamma_{2,k+1})+\sqrt{\Delta_{\gamma_{1,k}}}-\sqrt{\Delta_{\gamma_{1,k+1}}}<0
\end{eqnarray}
for $\Delta_{\gamma_{1,k}}$ and $\Delta_{\gamma_{1,k+1}}$ defined in (\ref{deltagamma1k}).
Indeed, inequality (\ref{gamma3kmon}) holds true since
$\sqrt{\Delta_{\gamma_{1,k}}}-\sqrt{\Delta_{\gamma_{1,k+1}}}<0$ which is equivalent to the inequality
\begin{eqnarray*}\lefteqn{(c_1-c_2)^2(\gamma_{2,k}-\gamma_{2,k+1})(\gamma_{2,k}+\gamma_{2,k+1})}\\&&+2(\gamma_{2,k}-\gamma_{2,k+1})(c_1-c_2)(\alpha c_1+\lambda+q)<0.\end{eqnarray*}
The latter inequality  follows from (\ref{gammapositive diff}). In the same way we can prove inequality $\gamma_{3,k+1}^\prime<\gamma_{3,k}^\prime$.
\exit

We will also  need some additional properties of the quantities introduced above which are collected in the next lemma.
\begin{lem}\label{properties} Let $k\in\N\cup\{0\}$. Then
\begin{description}
\item{(i)} 		$\lim_{k \rightarrow \infty} \gamma_{2,k}=\lim_{k \rightarrow \infty} \gamma_{2,k}^\prime=\infty$ and $\lim_{k \rightarrow \infty} \gamma_{i,k}=\lim_{k \rightarrow \infty} \gamma_{i,k}^\prime=-\infty$  for $i=1,3$;
\item{(ii)}		$\lim_{k \rightarrow \infty} \frac{\gamma_{i,k+1}}{\gamma_{i,k}}=\lim_{k \rightarrow \infty} \frac{\gamma_{i,k+1}^\prime}{\gamma_{i,k}^\prime}=\frac{c_1a+c_1}{c_1a+c_2}>1$ for $i=1,2,3$;
\item{(iii)}  $\lim_{k \rightarrow \infty} \frac{\gamma_{3,k}}{\gamma_{2,k}}=\lim_{k \rightarrow \infty} \frac{\gamma_{3,k}^\prime}{\gamma_{2,k}^\prime}=-1$ and $\lim_{k \rightarrow \infty} \frac{\gamma_{1,k}}{\gamma_{2,k}}=\lim_{k \rightarrow \infty} \frac{\gamma_{1,k}^\prime}{\gamma_{2,k}^\prime}=$~$-\frac{c_2}{c_1}$.
\end{description}
\end{lem}
\proof
Recall that by Lemma \ref{recc}(iii) the sequence $\{\gamma_{2,k}, k=0,1,2,\dots\}$ is monotone.
Hence if it were not true that $\lim_{k \rightarrow \infty} \gamma_{2,k}=\infty$, there would exist $C>0$ such that
$C=\lim_{k \rightarrow \infty} \gamma_{2,k}$. Then by (\ref{gammapositive diff}),
\begin{eqnarray}\nonumber
\lim_{k \rightarrow \infty} (\gamma_{2,k+1}-\gamma_{2,k})
&=& \lim_{k \rightarrow \infty}\left( \sqrt{\Delta_{\gamma_{2,k+1}}} + (2c_1a+c_1+c_2 )\sqrt{\Delta_{\gamma_{1,k}}}\right) \\
\nonumber &&+ \frac{(c_1-c_2)^2}{2c_1}C + \frac{(c_1-c_2)}{2}(\alpha +\lambda+q)>0.
\end{eqnarray}
But if $\lim_{k \rightarrow \infty} \gamma_{2,k}=C$ for some constant $C$ then $\lim_{k \rightarrow \infty} (\gamma_{2,k+1}-\gamma_{2,k})=0$ which gives a contradiction.
Note that by (\ref{gama3}) we have that $\lim_{k \rightarrow \infty}\frac{\gamma_{3,k}}{\gamma_{2,k}}= -1$ and hence
$\lim_{k \rightarrow \infty}\gamma_{3,k}= -\infty$ and
by (\ref{gammapositive diff}),
\begin{eqnarray}\lim_{k \rightarrow \infty}\frac{\gamma_{2,k+1}}{\gamma_{2,k}}= \frac{ac_1+c_1}{ac_1+c_2}>1.
\end{eqnarray}
Similarly, by (\ref{gama1}) $\lim_{k \rightarrow \infty}\frac{\gamma_{1,k}}{\gamma_{2,k}}\rightarrow -\frac{c_2}{c_1}$, hence
$\lim_{k \rightarrow \infty}\gamma_{1,k}= -\infty$.
Moreover, $\lim_{k \rightarrow \infty}\frac{\gamma_{1,k+1}}{\gamma_{2,k}}= -\frac{c_2}{c_1} \frac{ac_1+c_1}{ac_1+c_2} $ and thus
$\lim_{k \rightarrow \infty} \frac{\gamma_{1,k+1}}{\gamma_{1,k}}=\frac{c_1a+c_1}{c_1a+c_2}$. Similarly we can prove that
$\lim_{k \rightarrow \infty} \frac{\gamma_{3,k+1}}{\gamma_{3,k}}=\frac{c_1a+c_1}{c_1a+c_2}$.
The same limits we can obtain by exchanging $\gamma_{i,k}$ by $\gamma_{i,k}^\prime$ for $i=1,2,3$.
\exit

We now introduce recursively the following coefficients:
\begin{equation}\label{Lemma31ii}
D_0=e^{-\gamma_{2,0} b} \frac{(c_1+1) +(c_2-a)}{\gamma_{1,0}(c_1+1) +\gamma_{2,0}(a)(c_2-a)};
\end{equation}
\begin{eqnarray}
D_{k+1}=\frac{\gamma_{3,k}+\gamma_{2,k}+\alpha}{\gamma_{1,k}+\gamma_{2,k}+\alpha} \cdot \frac{\gamma_{3,k}(c_1+1)+\gamma_{2,k}(c_2-a)}{
\gamma_{1,k+1}(c_1+1)+\gamma_{2,k+1}(c_2-a)}D_k e^{\gamma_{2,k} b} e^{-\gamma_{2,k+1} b}.\label{Lemma31vi}
\end{eqnarray}
Similarly, $D_0^\prime=1$ and
\begin{eqnarray}
D_{k+1}^\prime=\frac{\gamma_{3,k}^\prime+\gamma_{2,k}^\prime+\alpha}{\gamma_{1,k}^\prime+\gamma_{2,k}^\prime+\alpha} \cdot \frac{\gamma_{3,k}^\prime(c_1+1)+\gamma_{2,k}^\prime(c_2-a)}{
\gamma_{1,k+1}^\prime(c_1+1)+\gamma_{2,k+1}^\prime(c_2-a)}D_k^\prime e^{\gamma_{2,k}^\prime b} e^{-\gamma_{2,k+1}^\prime b}.\label{Lemma31viprime}
\end{eqnarray}

The main result of this section gives a representation of the value function when claims are exponentially distributed with parameter $\alpha$ and the controlled process is reflected at the line
$z=b-ax$.
\begin{Thm}
For $(u_1,u_2)\in \mathcal{B}^{\rm c}$ we have:
\begin{eqnarray}
V_1(u_1,u_2)&=&\sum_{k=0}^{\infty} D_k \left( e^{\gamma_{1,k} u_1} -\frac{\gamma_{3,k}+\gamma_{2,k}+\alpha}{\gamma_{1,k}+\gamma_{2,k}+\alpha}e^{\gamma_{3,k} u_1}\right)e^{\gamma_{2,k} u_2}\label{2.7}\\
&&+E\sum_{k=0}^{\infty} D_k^\prime \left( e^{\gamma_{1,k}^\prime u_1} -\frac{\gamma_{3,k}^\prime+\gamma_{2,k}^\prime+\alpha}{\gamma_{1,k}^\prime+\gamma_{2,k}^\prime+\alpha}e^{\gamma_{3,k}^\prime u_1}\right)e^{\gamma_{2,k}^\prime u_2},\label{2.7prime}
\end{eqnarray}
where
\begin{equation}\label{E}
E=-\left.\left(\sum_{k=0}^{\infty} D_k \frac{\gamma_{1,k}-\gamma_{3,k}}{\gamma_{1,k}+\gamma_{2,k}+\alpha}e^{\gamma_{2,k}b}\right)\middle/ \left(\sum_{k=0}^{\infty} D_k^\prime \frac{\gamma_{1,k}^\prime-\gamma_{3,k}^\prime}{\gamma_{1,k}^\prime+\gamma_{2,k}^\prime+\alpha}e^{\gamma_{2,k}^\prime b}\right)\right. .
\end{equation}
\end{Thm}
\proof
In the first step of the proof we show the convergence of the series (\ref{2.7}) and (\ref{2.7prime}).
To do that we will use the d'Alembert criterion.
To prove that (\ref{2.7}) is convergent is sufficient to prove that the series
$\sum_{k=0}^{\infty} D_k e^{\gamma_{2,k} u_2} e^{\gamma_{1,k} u_1}$ and $\sum_{k=0}^{\infty} D_k e^{\gamma_{2,k} u_2}\frac{\gamma_{3,k}+\gamma_{2,k}+\alpha}{\gamma_{1,k}+\gamma_{2,k}+\alpha}e^{\gamma_{3,k} u_1}$ converge.
Note that:
\begin{eqnarray}\nonumber
\nonumber \lefteqn{\frac{D_{k+1}}{D_k} e^{\gamma_{2,k+1}u_2}
e^{-\gamma_{2,k}u_2} e^{\gamma_{1,k+1}u_1} e^{-\gamma_{1,k}u_1}}\\&&=
\nonumber  \frac{\gamma_{3,k}+\gamma_{2,k}+\alpha}{\gamma_{1,k}+\gamma_{2,k}+\alpha}\cdot \frac{\gamma_{3,k}(c_1+1)+\gamma_{2,k}(c_2-a)}{
\gamma_{1,k+1}(c_1+1)+\gamma_{2,k+1}(c_2-a)}\\&& \qquad\qquad e^{(\gamma_{2,k+1}-\gamma_{2,k})(u_2- b)} e^{(\gamma_{1,k+1}-\gamma_{1,k})u_1}
\nonumber \rightarrow 0\qquad\mbox{as $k\to\infty$.}
\end{eqnarray}
The last limit statement follows from Lemma \ref{properties} and the observation that for $b>u_2$,
$$ \lim_{k \rightarrow \infty}\frac{\gamma_{3,k}+\gamma_{2,k}+\alpha}{\gamma_{1,k}+\gamma_{2,k}+\alpha} =
\lim_{k \rightarrow \infty}e^{(\gamma_{2,k+1}-\gamma_{2,k})(u_2- b)}=\lim_{k \rightarrow \infty}e^{(\gamma_{1,k+1}-\gamma_{1,k})u_1}=0
$$
and
$$\lim_{k \rightarrow \infty}\frac{\gamma_{3,k}(c_1+1)+\gamma_{2,k}(c_2-a)}{
\gamma_{1,k+1}(c_1+1)+\gamma_{2,k+1}(c_2-a)} = \textrm{ const. }$$
Similarly,
\begin{eqnarray}\nonumber
\nonumber \lefteqn{\frac{D_{k+1}}{D_k} e^{\gamma_{2,k+1}u_2}
e^{-\gamma_{2,k}u_2} e^{\gamma_{3,k+1}u_1} e^{-\gamma_{3,k}u_1}} \cdot \\&& \frac{\gamma_{3,k+1}+\gamma_{2,k+1}+\alpha}{\gamma_{1,k+1}+\gamma_{2,k+1}+\alpha} \cdot \frac{\gamma_{1,k}+\gamma_{2,k}+\alpha}{\gamma_{3,k}+\gamma_{2,k}+\alpha} \\&&=
\nonumber  \frac{\gamma_{3,k+1}+\gamma_{2,k+1}+\alpha}{\gamma_{1,k+1}+\gamma_{2,k+1}+\alpha} \cdot \frac{\gamma_{3,k}(c_1+1)+\gamma_{2,k}(c_2-a)}{
\gamma_{1,k+1}(c_1+1)+\gamma_{2,k+1}(c_2-a)}\\&& \qquad\qquad  e^{(\gamma_{2,k+1}-\gamma_{2,k})(u_2- b)} e^{(\gamma_{3,k+1}-\gamma_{3,k})u_1} \rightarrow 0\qquad\mbox{as $k\to\infty$.}
\nonumber
\end{eqnarray}
Convergence of the series appearing in (\ref{2.7prime}) we can prove in the same way.

In the next step we prove that the function $V_1:=V$ indeed solves equation (\ref{1.1}) which can be rewritten in the following way:
\begin{eqnarray}\label{2.0}
c_1 \frac{\partial V}{\partial u_1}+c_2 \frac{\partial V}{\partial u_2}-(\lambda +q) V(\underline{u})
 +\lambda \int_0^{\min(u_1,u_2)} V(\underline{u}-(1,1)v)\alpha e^{-\alpha v}  \, dv =0.
\end{eqnarray}

We will find the solution of equation (\ref{2.0})
from the class  $V\in \mathcal{C}^2(\mathbb{R}^2)$. The uniqueness of the solution of (\ref{2.0}) proved in Theorem \ref{Vneq}
ensures us that this solution is a proper solution.
If $V\in \mathcal{C}^2(\mathbb{R}^2)$ then applying an operator $\frac{\partial}{\partial u_1}+\frac{\partial}{\partial u_2}$ to the equation (\ref{2.0}) gives
\begin{eqnarray}
\nonumber c_1 \frac{\partial V}{\partial u_1^2}+c_2 \frac{\partial V}{\partial u_2^2}+(c_1+c_2) \frac{\partial V}{\partial u_1 \partial u_2}+(\alpha c_1 -q -\lambda)\frac{\partial }{\partial u_1}V(\underline{u})\\
+(\alpha c_2 -q-\lambda)\frac{\partial }{\partial u_2}V(\underline{u})-\alpha q V(\underline{u}) =0\label{2.1}
\end{eqnarray}
since
\begin{eqnarray}
\nonumber\lefteqn{\left(\frac{\partial}{\partial u_1}+\frac{\partial}{\partial u_2}\right)\int_0^{\min(u_1,u_2)} V(u_1-v,u_2-v) \alpha e^{-\alpha v} \,dv} \\ \nonumber&&
=-\alpha \int_0^{\min(u_1,u_2)} V(u_1-v,u_2-v)\alpha e^{-\alpha v}  \, dv +\alpha V(\underline{u}).
\end{eqnarray}
Since $V_0(\underline{u})=1$ the boundary condition (\ref{boundarythm}) translates into
\begin{eqnarray}\label{bound}
  (c_1+1)+(c_2-a)=\left.(c_1+1)\frac{\partial V}{\partial u_1}\right|_{\underline{u}\in\mathcal{B}_0}+\left.(c_2-a)\frac{\partial V}{\partial u_2}\right|_{\underline{u}\in\mathcal{B}_0},
\end{eqnarray}
where $\mathcal{B}_0$ is defined in (\ref{bar0}).

We will look first for a solution of equation (\ref{2.1}) of the following form:
\begin{eqnarray}\label{2.2}
\left(\Upsilon_1 e^{\gamma_1 u_1} +\Upsilon_2e^{\gamma_3 u_1}\right)e^{\gamma_2 u_2}
\end{eqnarray}
for some constants $\Upsilon_1$ and $\Upsilon_2$.
Putting (\ref{2.2}) into (\ref{2.1}) shows that $\gamma_1$ and $\gamma_3$ are two real roots of the equation:
\begin{eqnarray}\label{2.3}
c_1  \gamma^2 + c_2  \gamma_2^2 +(c_1+c_2) \gamma  \gamma_2 + (\alpha c_1 -q -\lambda)\gamma+
(\alpha c_2 -q-\lambda)\gamma_2 -\alpha q   =0.
\end{eqnarray}
Recall that $u_1<u_2$. Then from (\ref{2.0}) we obtain:
\begin{eqnarray}\label{2.4}
\lefteqn{ \Upsilon_1\left(c_1 \gamma_1 +c_2 \gamma_2 -(\lambda +\delta) + \frac{\lambda \alpha}{\gamma_1 + \gamma_2 +\alpha}\right)e^{\gamma_1 u_1 +\gamma_2 u_2}
}\nonumber\\
\nonumber&& -\frac{\lambda \Upsilon_1 \alpha}{\gamma_1+\gamma_2+\alpha}e^{-(\gamma_2 +\alpha )u_1 +\gamma_2 u_2}+\Upsilon_2\Bigl(c_1 \gamma_3 +c_2 \gamma_2 -(\lambda +\delta)
\Bigr.\\
\nonumber&&+\Bigl.\frac{\lambda \alpha}{\gamma_3 +\gamma_2 +\alpha}\Bigr)e^{\gamma_3 u_1 +\gamma_2 u_2}-\frac{\lambda \Upsilon_2 \alpha}{\gamma_3+\gamma_2+\alpha}e^{-(\gamma_2 +\alpha ) u_1 +\gamma_2 u_2}=0.
\end{eqnarray}
Note that $c_1 \gamma_1 +c_2 \gamma_2 -(\lambda +\delta) + \frac{\lambda \alpha}{\gamma_1 + \gamma_2 +\alpha}=c_1 \gamma_3 +c_2 \gamma_2 -(\lambda +\delta)+\frac{\lambda \alpha}{\gamma_3 +\gamma_2 +\alpha}=0$. Thus
$$ \Upsilon_2=\frac{-\Upsilon_1(\gamma_3+\gamma_2+\alpha)}{\gamma_1+\gamma_2+\alpha}$$
and (\ref{2.2}) can be rewritten in the following form:
\begin{eqnarray}\label{2.5}
\Upsilon_1\left( e^{\gamma_1 u_1} -\frac{\gamma_3+\gamma_2+\alpha}{\gamma_1+\gamma_2+\alpha}e^{\gamma_3 u_1}\right)e^{\gamma_2 u_2}
\end{eqnarray}

Comparing (\ref{2.3}) and (\ref{sqeq})-(\ref{sqeqprime}) one can conclude that
the function $V$ given in (\ref{2.7})-(\ref{2.7prime}) is a linear combination of functions of type (\ref{2.5}).
Hence it solves equation (\ref{2.0}).

We choose coefficients $\gamma_{2,0}$, $D_{k}$ and $D^\prime_k$ given in (\ref{gamma2zero}), (\ref{Lemma31ii}), (\ref{Lemma31vi}) and
(\ref{Lemma31viprime}) in such a way to satisfy the boundary condition (\ref{bound}) for $V_1$:
\begin{eqnarray}\label{2.9}
\nonumber \lefteqn{\sum_{k=0}^{\infty} D_k e^{\gamma_{2,k} (b-au_1)}\Bigl( e^{\gamma_{1,k} u_1} (\gamma_{1,k}(c_1+1)+\gamma_{2,k}(c_2-a))\Bigr.}
\\ \nonumber &&\Bigl. -\frac{\gamma_{3,k}+\gamma_{2,k}+\alpha}{\gamma_{1,k}+\gamma_{2,k}+\alpha}
(\gamma_{3,k}(c_1+1)+\gamma_{2,k}(c_2-a)) e^{\gamma_{3,k} u_1}\Bigr)\\&&=(c_1+1) +(c_2-a)
\end{eqnarray}
and
\begin{eqnarray}\label{2.9prime}
\nonumber \lefteqn{\sum_{k=0}^{\infty} D_k^\prime e^{\gamma_{2,k}^\prime (b-au_1)}\Bigl( e^{\gamma_{1,k}^\prime u_1} (\gamma_{1,k}^\prime(c_1+1)+\gamma_{2,k}^\prime(c_2-a))\Bigr.}
\\ \nonumber &&\Bigl. -\frac{\gamma_{3,k}^\prime+\gamma_{2,k}^\prime+\alpha}{\gamma_{1,k}^\prime+\gamma_{2,k}^\prime+\alpha}
(\gamma_{3,k}^\prime(c_1+1)+\gamma_{2,k}^\prime(c_2-a)) e^{\gamma_{3,k}^\prime u_1}\Bigr)\\&&=0.
\end{eqnarray}
Note that (\ref{uciekabariera}) holds also true since $\gamma_{2,k}>0$ and $\gamma_{2,k}^\prime>0$ ($k=0,1,\ldots$)
by Lemma \ref{recc}.
The coefficient $E$ is chosen in such way to satisfy last boundary condition (\ref{gornyrog}).
Note also that there will be no more functions
being the linear combination of functions of type (\ref{2.5}) such that (\ref{2.9}) and (\ref{2.9prime}) are satisfied
since all coefficients of $V$ are uniquely determined.
This observation completes the proof.
\exit

For the numerical analysis we assume $\alpha=2$, $c_1=4$, $c_2=3$, $\lambda=1$, $q=0.1$. The values of expected dividend payments $V_1(u_1,u_2)$ for $u_1=1$, $u_2=2$  and $u_1=2$, $u_2=3$ depending on $a$ and $b$ are given in the Tables 1 and 2
below. Note that there always exists an optimal choice of linear barrier (choice of its upper left end $(0,b)$ and it slope $a$).
This choice depends on the initial reserves $(u_1,u_2)$. For $(u_1,u_2)=(1,2)$ the optimal barrier is determined by $b=14$ and $a=0.1$ and for $(u_1,u_2)=(2,3)$ the optimal barrier is determined by $b=15$ and $a=0.1$. This is in contrast with the one-dimensional case where the choice of the barrier is given only via the premium rate and the distribution of the arriving claims.

\begin{center}
\begin{tabular}{|c||c|c|c|c|c|c|}
\hline
  &\multicolumn{6}{c|}{$b$}\\
\cline{2-7}
$a$ & 6 & 8 & 14& 15 & 20 & 28 \\
\hline\hline
0.1 & 19.85 & 27.20 & 34.95& 34.93 & 32.48 & 25.89  \\
0.2 & 16.33&  24.31 & 33.82& 34.19 & 33.32 & 28.03  \\
0.5 & 11.76& 17.74 &28.98 & 30.01 & 32.54 & 31.21  \\
1 & 7.22 & 11.40 & 21.35& 22.59 & 27.17 & 30.07 \\
\hline
\end{tabular}
\vspace{0.5cm}

Table 1: Expected value of dividend payments depending on $a$ and $b$ for fixed $(u_1,u_2)=(1,2)$.
\end{center}

\begin{center}
\begin{tabular}{|c||c|c|c|c|c|c|}
\hline
  &\multicolumn{6}{c|}{$b$}\\
\cline{2-7}
$a$ & 6 & 8 & 14& 15 & 20 & 28 \\
\hline\hline
0.1 & 19.07 & 27.42&36.51 & 36.58 & 34.21 & 27.34  \\
0.2 & 17.17&  24.34 & 35.22& 35.69& 35.01 & 29.55  \\
0.5 & 10.94& 17.50 & 29.93& 31.07 & 33.99 & 32.78  \\
1 & 6.59 & 11.07 &21.86 & 23.21 & 28.19 & 31.43 \\
\hline
\end{tabular}
\vspace{0.5cm}

Table 2: Expected value of dividend payments depending on $a$ and $b$ for fixed $(u_1,u_2)=(2,3)$.
\end{center}

The Table 3 gives $V(u_1,u_2)$ for fixed $a=0.9$ and $b=1.8$ when $u_1<u_2$ and $u_2\leq b-au_1$.
From this table for example if follows that for a given slope $a$ of the linear barrier
it is optimal to locate an initial capital around some line. In this case it is a line $u_2=u_1$.

\begin{center}
\begin{tabular}{|c||c|c|c|c|c|c|}
\hline
  &\multicolumn{6}{c|}{$u_2$}\\
\cline{2-7}
$u_1$ & 0.2& 0.4 &0.6 &0.8 &0.9& 1.2 \\
\hline\hline
0    & 2.09 & 1.58 & 1.11 &0.69 & 0.49 & 0.03  \\
0.1  & 2.35 & 1.81 & 1.31 & 0.86 & 0.65 & 0.13   \\
0.2  &  & 2.06 & 1.53 & 1.09 & 0.82 & 0.25  \\
0.4  &  &  & 1.98 & 1.45 & 1.20 & 0.53  \\
0.7  &  & &  & 2.11 & 1.83 & \\
0.8	  &  &  &  &  & 2.07 & \\
\hline
\end{tabular}
\vspace{0.5cm}

Table 3: Expected value of dividend payments depending on $u_1$ and $u_2$ for fixed $a$ and $b$.
\end{center}

\sec{The impulse control scheme}\label{sec:second}

In this section we consider the impulse controlling. The size of the $i$th payment made at the epoch $T_i$
equals $J_i=X_1(T_i)-u_1+(c_1+c_2)e_\lambda^{(i)}$, where $T_i$ is the $i$th moment when $\underline{X}(t)$ hits the horizontal line $z=u_2$ and
$e^{(i)}_\lambda$ is independent of $X$ the exponential random variable
with intensity $\lambda$. It means that the controlled risk process always starts at $(u_1,u_2)$. Then until the first claim arrives the premium is transferred into dividend payments and controlled process
stays at $(u_1, u_2)$. Just right after of arrival of the first claim the controlled process evolves without paying dividends until it hits horizontal line $z=u_2$.
Then each payment of the dividend corresponds to reducing reserves to some fixed levels $(u_1,u_2)$ (see Figure 2) and paying out dividends while
waiting for the next claim to arrive. By (\ref{c12}) at the time $T_i$
the first company has more reserves than $u_1$. Hence the impulse payments $X_1(T_i)-u_1$
of the dividends are always made by the first company which has greater premium rate (by reducing the reserves to
the level $u_1$).
Associated to each dividend payment is a fixed
cost of size $K>0$.

We will consider two cases, when $u_1 > u_2$ and when $u_1\leq u_2$. In the first case
the ruin can be only achieved by the second insurance company in contrast to the second case when both companies might get ruined.

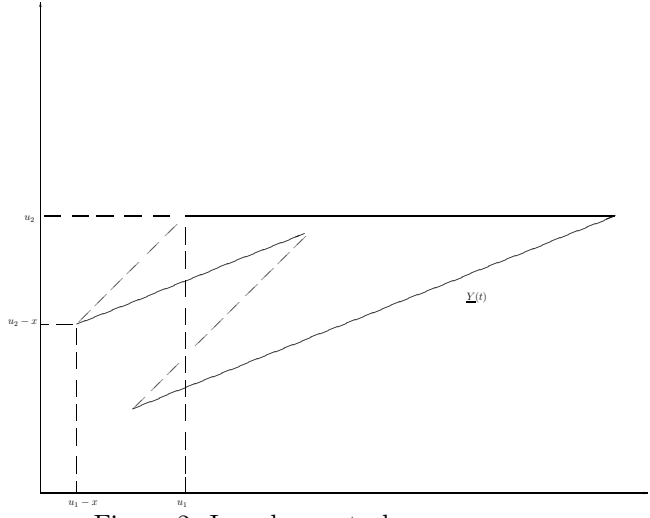
\begin{figure}[t!] \label{piece}\centering
\resizebox{.34 \textwidth}{!}{
\centerline{\setlength{\unitlength}{3947sp}%
\begingroup\makeatletter\ifx\SetFigFont\undefined%
\gdef\SetFigFont#1#2#3#4#5{%
  \reset@font\fontsize{#1}{#2pt}%
  \fontfamily{#3}\fontseries{#4}\fontshape{#5}%
  \selectfont}%
\fi\endgroup%
\begin{picture}
(5000, 4000)(2000,-8502)
\thinlines
{\put(601,-8836){\vector( 0, 1){9150}}
}%
{\put(601,-8836){\vector( 1, 0){11325}}
}%

{\put(3001,-3661){\line(-1, 0){300}}
}%

{\put(2476,-3661){\line(-1, 0){300}}
}%
{\put(1951,-3661){\line(-1, 0){300}}
}%
{\put(1501,-3661){\line(-1, 0){300}}
}%
{\put(976,-3661){\line(-1, 0){300}}
}%
{\put(3751,-3661){\line(-1, 0){300}}
}%

{\put(3301,-3886){\line( 0,-1){300}}
}%
{\put(3301,-4336){\line( 0,-1){300}}
}%
{\put(3301,-4861){\line( 0,-1){300}}
}%
{\put(3301,-5386){\line( 0,-1){300}}
}%
{\put(3301,-5836){\line( 0,-1){300}}
}%
{\put(3301,-6286){\line( 0,-1){300}}
}%
{\put(3301,-6736){\line( 0,-1){300}}
}%
{\put(3301,-7186){\line( 0,-1){300}}
}%
{\put(3301,-7711){\line( 0,-1){300}}
}%
{\put(3301,-8161){\line( 0,-1){300}}
}%
{\put(3301,-8536){\line( 0,-1){300}}
}%
{\put(3151,-3811){\line(-1,-1){225}}
}%
{\put(2813,-4148){\line(-1,-1){225}}
}%
{\put(2438,-4523){\line(-1,-1){225}}
}%
{\put(2138,-4823){\line(-1,-1){225}}
}%
{\put(1838,-5123){\line(-1,-1){225}}
}%
{\put(1538,-5423){\line(-1,-1){225}}
}%
\thicklines

{\put(1276,-5686){\line( 5, 2){4254.310}}
}%

\thinlines
{\put(5551,-4036){\line(-1,-1){225}}
}%
{\put(5251,-4336){\line(-1,-1){225}}
}%
{\put(4951,-4636){\line(-1,-1){225}}
}%
{\put(4651,-4936){\line(-1,-1){225}}
}%
{\put(4313,-5273){\line(-1,-1){225}}
}%
{\put(4013,-5573){\line(-1,-1){225}}
}%
{\put(3713,-5873){\line(-1,-1){225}}
}%
{\put(3188,-6398){\line(-1,-1){225}}
}%
{\put(2888,-6698){\line(-1,-1){225}}
}%
{\put(2551,-7036){\line(-1,-1){225}}
}%
\thicklines

{\put(2325,-7258){\line( 5, 2){
9000}}
}%

{\put(3301,-3661){\line( 1, 0){6150}}
}%
{\put(9451,-3661){\line( 1, 0){1850}}
}%
\thinlines
{\put(1276,-5761){\line( 0,-1){300}}
}%
{\put(1276,-6211){\line( 0,-1){300}}
}%
{\put(1276,-6736){\line( 0,-1){300}}
}%
{\put(1276,-7186){\line( 0,-1){300}}
}%
{\put(1276,-7636){\line( 0,-1){300}}
}%
{\put(1276,-8161){\line( 0,-1){300}}
}%
{\put(1276,-8611){\line( 0,-1){225}}
}%
{\put(1201,-5686){\line(-1, 0){300}}
}%
{\put(751,-5686){\line(-1, 0){150}}
}%
\put(301,-3736){\makebox(0,0)[lb]{\smash{{\SetFigFont{12}{14.4}{\rmdefault}{\mddefault}{\updefault}{$u_2$}%
}}}}
\put(3151,-9061){\makebox(0,0)[lb]{\smash{{\SetFigFont{12}{14.4}{\rmdefault}{\mddefault}{\updefault}{$u_1$}%
}}}}
\put(1126,-9061){\makebox(0,0)[lb]{\smash{{\SetFigFont{12}{14.4}{\rmdefault}{\mddefault}{\updefault}{$u_1-x$}%
}}}}
\put(
0,-5686){\makebox(0,0)[lb]{\smash{{\SetFigFont{12}{14.4}{\rmdefault}{\mddefault}{\updefault}{$u_2-x$}%
}}}}
\put(8551,-5236){\makebox(0,0)[lb]{\smash{{\SetFigFont{12}{14.4}{\rmdefault}{\mddefault}{\updefault}{$\underline{Y}(t)$}%
}}}}
\end{picture}%
}
}

\caption{Impulse control. }
\end{figure}

\subsection{Case $u_1>u_2$}
Define: \begin{eqnarray*}&&\tau_{x}^{+}=\inf\{t\geq 0: c_2t-S(t)=x\},\\&&
\tau_{-u_2+x}^{-}=\inf\{t \geq 0: c_2t-S(t)<-(u_2-x)\}.
\end{eqnarray*}
Let $x$ be the size of the first claim $U_1$ chosen according to the density function $f$.
If there is no ruin at the moment of the arrival of the first claim then $x\leq u_2$.
When the first claim arrives the risk process jumps
from $(u_1,u_2)$ to $(u_1-x, u_2-x)$.
Note that the first portion $A$ of payed dividends equals the discounted independent exponential random variable $e^{(1)}_\lambda=e_\lambda$ with the parameter $\lambda$
multiplied by $c_1+c_2$
plus discounted jump on the line $y=u_2$ which equals $(c_1-c_2)\tau_{x}^{+}$ minus the costs of one payment made at the moment of the impulse
payment (that is at the moment of jump at the line $y=u_2$). Thus,
 \begin{eqnarray}\nonumber
\nonumber \lefteqn{A:=(c_1+c_2)\Exp\left[\int_0^{e_{\lambda}}e^{-qt}\, dt\right]}\\&&
\nonumber\quad + \int_{0}^{u_2} \Exp\left[\left((c_1-c_2)\tau_{x}^{+}-K\right) e^{-q (\tau_{x}^{+}+e_{\lambda})}
\textrm{ } 1_{(\tau_{x}^{+}<\tau_{-u_2+x}^{-})}\right] f(x)\, dx\\&&
\nonumber =\frac{c_1+c_2}{q+\lambda}
\nonumber +\frac{\lambda}{q+\lambda} \int_0^{u_2}\Exp\left[\left((c_1-c_2)\tau_{x}^{+}-K\right) e^{-q \tau_{x}^{+}}
\textrm{ } 1_{(\tau_{x}^{+}<\tau_{-u_2+x}^{-})}\right]f(x) \, dx.
\end{eqnarray}
Then the mean of the cumulative discounted dividends equals:
\begin{eqnarray}\label{1.5}
\nonumber \lefteqn{V_1(u_1,u_2)=A+ A \underbrace{ \int_{0}^{u_2} \Exp[ e^{-q (e_{\lambda}+\tau_{x}^{+})} \textrm{   }  1_{(\tau_{x}^{+}<\tau_{-u_2+x}^{-})}] f(x)\, dx}_{p}}\\&&
 + A p^2+A p^3  +...= A\sum_{i=1}^{\infty} p^{i-1} = \frac{A}{1-p},
\end{eqnarray}
where
$$p= \frac{\lambda}{q+\lambda} \int_{0}^{u_2}\Exp[e^{-q \tau_{x}^{+}}
\textrm{ } 1_{(\tau_{x}^{+}<\tau_{-u_2+x}^{-})}] f(x)\, dx.$$
From Kyprianou \cite[Th.8.1, p. 214]{Kbook} it follows that
\begin{eqnarray}\nonumber
&&\Exp[e^{-q  \tau_{x}^{+}} \textrm{   } 1_{( \tau_{x}^{+}<\tau_{-u_2+x}^{-})}]=\frac{W^{(q)}(u_2-x)}{W^{(q)}(u_2)},
\end{eqnarray}
where $W^{(q)}: [0,\i) \to [0,\i)$ is a scale function, that is continuous and
increasing function with the Laplace transform \eqn{\label{eq:defW} \I_0^\i
\te{-\th x} W^{(q)} (y)  \td y = (\ps(\th) - q)^{-1},\q\q\th >
\F(q), } for $\psi(\th)=c_2\theta +\lambda(Ee^{-\theta U}-1)$ being the Laplace exponent of $X_1$ and for $\F(q)=\sup\{\th \geq 0: \psi(\th)=q\}$
being its right inverse. Moment $\tau_x^+1_{(\tau_{x}^{+}<\tau_{-u_2+x}^{-})}$ has continuous density which follows from Kendall's formula applied to Feller process $c_2t-S(t)$
killed when it enters into the halfline $(-\infty, -u_2+x)$.
Moreover, $\Exp[\tau_{x}^{+}e^{-q  \tau_{x}^{+}} \textrm{   } 1_{( \tau_{x}^{+}<\tau_{-u_2+x}^{-})}]\leq \E\tau_x^+=\frac{x}{c_2-\E U}$, where the last equality is a consequence of Wald identity.
Hence one can take the derivative under the integral sign with respect to parameter $q$ giving:
\begin{eqnarray}\label{pochzam}
&&\Exp[\tau_{x}^{+}e^{-q  \tau_{x}^{+}} \textrm{   } 1_{( \tau_{x}^{+}<\tau_{-u_2+x}^{-})}]
=-\frac{d}{dq}\left( \frac{W^{(q)}(u_2-x)}{W^{(q)}(u_2)}\right).
\end{eqnarray}

For the exponentially distributed claim sizes with parameter $\alpha$
we have that
$\psi(\theta)=c_2\theta-\lambda \theta/(\alpha+\theta)$
and the scale function $W^{(q)}$ is given by
$$W^{(q)}(x) = c_2^{-1}\left(A_+ e^{q^+(q)x} - A_-e^{q^-(q)x}\right),$$
where $A_\pm = \frac{\alpha + q^\pm(q)}{q^+(q)-q^-(q)}$ and
$$q^\pm (q)=\frac{q + \lambda -\alpha c_2 \pm \sqrt{(q + \lambda-\alpha c_2)^2 + 4 c_2 q\alpha}}{2 c_2}.$$
Then,
\begin{eqnarray}\nonumber
p&=&
\frac{\lambda \alpha}{c_2(q+\lambda)W^{(q)}(u_2)}\cdot \frac{e^{q^+(q)u_2}-e^{q^-(q)u_2}}{q^+(q)-q^-(q)}.
\end{eqnarray}
and
$$A=\frac{c_1+c_2}{q+\lambda}-Kp+\frac{\lambda}{q+\lambda}(c_1-c_2)\int_{0}^{u_2} \Exp\left[\tau_{x}^{+} e^{-q \tau_{x}^{+}}
\textrm{ } 1_{(\tau_{x}^{+}<\tau_{-u_2+x}^{-})}\right] f(x)\, dx$$
with
\begin{eqnarray*}\nonumber
\lefteqn{\int_{0}^{u_2} \Exp\left[\tau_{x}^{+} e^{-q \tau_{x}^{+}}
\textrm{ } 1_{(\tau_{x}^{+}<\tau_{-u_2+x}^{-})}\right] f(x)\, dx} \\ \nonumber &&=\frac{\frac{d}{dq}W^{(q)}(u_2)}{(W^{(q)}(u_2))^2} \int_{0}^{u_2} W^{(q)}(u_2-x) \alpha e^{-\alpha x} \, dx \\ \nonumber &&\quad-
\frac{1}{W^{(q)}(u_2)} \int_{0}^{u_2} \frac{d}{dq}W^{(q)}(u_2-x) \alpha e^{-\alpha x} \, dx\\&&=
\frac{\frac{d}{dq}W^{(q)}(u_2)}{(W^{(q)}(u_2))^2}\frac{\alpha}{c_2}  \frac{e^{q^+(q)u_2}-e^{q^-(q)u_2}}{q^+(q)-q^-(q)}\\&&\quad+
\frac{1}{W^{(q)}(u_2)}\Bigl[\frac{\alpha}{c_2(q^+(q)+\alpha)}A'_+\left(e^{q^+(q)u_2}
-e^{-\alpha u_2}\right)\Bigr.\\&&\quad-\frac{\alpha}{c_2(q^-(q)+\alpha)}A'_-\left(e^{q^-(q)u_2}-e^{-\alpha u_2}\right)\\ \nonumber &&\quad+
 \frac{\alpha}{c_2(q^+(q)+\alpha)}q^+(q)'A_+\left(e^{q^+(q)u_2}-e^{-\alpha u_2}\right)\\&&\quad \Bigl.-\frac{\alpha}{c_2(q^-(q)+\alpha)}q^-(q)'A_-\left(e^{q^-(q)u_2}-e^{-\alpha u_2}\right)\Bigr],
\end{eqnarray*}
where
\begin{eqnarray*}&&q^\pm(q)'= \frac{1}{2c_2}\left[1\pm\frac{q+\lambda+\alpha c_2}{\sqrt{{(q + \lambda-\alpha c_2)^2 + 4 c_2 q\alpha}}}\right],\\&&
A'_\pm= \frac{d}{dq}A_\pm=\frac{q^\pm(q)'(q^+(q)-q^-(q))-(\alpha +q^\pm(q))(q^+(q)'-q^-(q)')}{(q^+(q)-q^-(q))^2},\\
&&\frac{d}{dq}W^{(q)}(x)=
c_2^{-1}\left(  A'_+ e^{q^+(q)x} + q^+(q)'A_+ e^{q^+(q)x} -  A'_- e^{q^-(q)x} - q^-(q)'A_- e^{q^-(q)x} \right).
\end{eqnarray*}

\subsection{Case $u_1\leq u_2$}

Let $x$ be the size of the first claim $U_1$ chosen according to the distribution function $F$.
If there is no ruin at the moment of the arrival of the first claim then $x\leq u_1$.
Define
$$\tau_{U}=\inf\{t\geq 0: Z(t)=u_2\},$$
$$\tau_{L}=\inf\{t \geq 0: Z(t)< \max (0,u_2-u_1-(c_1-c_2)t)\},$$
where $Z(t)=(u_2-x)+c_2 t -S(t)$.
Note that when $X_1(0)=u_1-x, X_2(0)=Z(0)=u_2-x\geq u_2-u_1$, then $\{X_1(t)\geq 0, X_2(t)\geq 0\}$ is equivalent to the requirement that
$\{\tau_L>t\}$ (see Figure 3).
\begin{figure}[t!] \label{piece}\centering
\resizebox{.34 \textwidth}{!}{
\centerline{\setlength{\unitlength}{3947sp}%
\begingroup\makeatletter\ifx\SetFigFont\undefined%
\gdef\SetFigFont#1#2#3#4#5{%
  \reset@font\fontsize{#1}{#2pt}%
  \fontfamily{#3}\fontseries{#4}\fontshape{#5}%
  \selectfont}%
\fi\endgroup%
\begin{picture}
(5000, 4000)(2000,-8502)
\thinlines
{\put(676,-8686){\vector( 0, 1){9150}}
}%
{\put(676,-8686){\vector( 1, 0){11850}}
}%
\thicklines
{\put(676,-4636){\line( 1,-1){4050}}
}%
{\put(676,-1411){\line( 1, 0){10425}}
}%
{\put(676,-3211){\line( 6, 5){
1593.443
}}
}%
\thinlines
{\put(2251,-1861){\line( 0,-1){375}}
}%
{\put(2251,-2311){\line( 0,-1){375}}
}%
{\put(2251,-2836){\line( 0,-1){375}}
}%
{\put(2251,-3361){\line( 0,-1){375}}
}%
{\put(2251,-3886){\line( 0,-1){375}}
}%
{\put(2251,-4411){\line( 0,-1){
375}}
}%
\thicklines
{\put(2256,-4792){\line( 6, 5){
5600
}}
}%
\thinlines
{\put(7876, -150){\line( 0,-1){
100
}}
}%
{\put(7876,-436){\line( 0,-1){375}}
}%
{\put(7876,-961){\line( 0,-1){375}}
}%
{\put(7876,-1561){\line( 0,-1){375}}
}%
{\put(7876,-2086){\line( 0,-1){375}}
}%
{\put(7876,-2611){\line( 0,-1){375}}
}%
{\put(7876,-3136){\line( 0,-1){375}}
}%
{\put(7876,-3661){\line( 0,-1){375}}
}%
{\put(7876,-4186){\line( 0,-1){375}}
}%
{\put(7876,-4786){\line( 0,-1){375}}
}%
{\put(7876,-5386){\line( 0,-1){375}}
}%
{\put(7876,-5986){\line( 0,-1){375}}
}%
\thicklines
{\put(7913,-6404){\line( 6, 5){
3100
}}
}
\thinlines
{\put(11026,-3811){\line( 0,-1){375}}
}%
{\put(11026,-4336){\line( 0,-1){375}}
}%
{\put(11026,-4861){\line( 0,-1){375}}
}%
{\put(11026,-5386){\line( 0,-1){375}}
}%
{\put(11026,-5911){\line( 0,-1){375}}
}%
{\put(11026,-6511){\line( 0,-1){375}}
}%
{\put(11026,-7111){\line( 0,-1){375}}
}%
{\put(11026,-7636){\line( 0,-1){375}}
}%
{\put(11026,-8236){\line( 0,-1){375}}
}%
\put(4651,-8911){\makebox(0,0)[lb]{\smash{{\SetFigFont{12}{14.4}{\rmdefault}{\mddefault}{\updefault}{$R$}%
}}}}
\put( 0,-4636){\makebox(0,0)[lb]{\smash{{\SetFigFont{12}{14.4}{\rmdefault}{\mddefault}{\updefault}{$u_2-u_1$}%
}}}}
\put(376,-1411){\makebox(0,0)[lb]{\smash{{\SetFigFont{12}{14.4}{\rmdefault}{\mddefault}{\updefault}{$u_2$}%
}}}}
\put(12301,-8911){\makebox(0,0)[lb]{\smash{{\SetFigFont{12}{14.4}{\rmdefault}{\mddefault}{\updefault}{$t$}%
}}}}
\put(3826,-2986){\makebox(0,0)[lb]{\smash{{\SetFigFont{12}{14.4}{\rmdefault}{\mddefault}{\updefault}{$Z(t)$}%
}}}}
\end{picture}%
}

} \caption{Piecewise lower barrier when $u_1\leq u_2$. }
\end{figure}
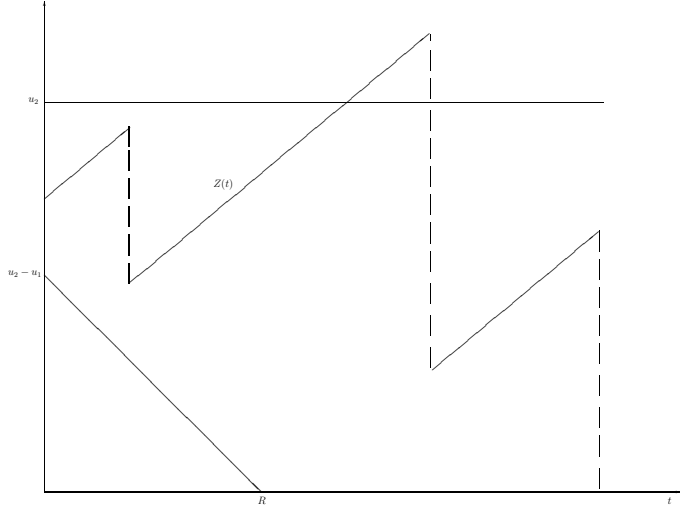
Moreover, the impulse payment equals
$(c_1-c_2)\tau_{U}.$
Therefore, like in the previous case:
\begin{eqnarray}\label{1.5b}
V_1(u_1,u_2)&=& \frac{A}{1-p},
\end{eqnarray}
where
$$p= \frac{\lambda}{q+\lambda} \int_{0}^{u_2}\Exp\left[e^{-q \tau_{U}}
\textrm{ } 1_{(\tau_{U}<\tau_{L})}\right]f(x) \, dx$$
and
\begin{eqnarray}\nonumber
\nonumber &&A:=
\frac{c_1+c_2}{q+\lambda}-Kp
+\frac{\lambda(c_1-c_2)}{q+\lambda}\int_{0}^{u_2} \Exp \left[\tau_{U} e^{-q \tau_{U}}
\textrm{ } 1_{(\tau_{U}<\tau_{L})}\right] f(x)\, dx
\end{eqnarray}
with
\begin{eqnarray}
\Exp\left[\tau_U e^{-q  \tau_{U}} \textrm{   } 1_{( \tau_{U}<\tau_{L})}\right]=
-\frac{d}{dq} \Exp\left[e^{-q  \tau_{U}} \textrm{   } 1_{( \tau_{U}<\tau_{L})}\right].\label{pochodnatau}
\end{eqnarray}
To compute $\Exp\left[e^{-q  \tau_{U}} \textrm{   } 1_{( \tau_{U}<\tau_{L})}\right]$ we introduce a new probability measure $\P^{\Phi(q)}$:
$$\left.\frac{d\P^{\Phi(q)}}{d\P}\right|_{\mathcal{F}_t}=e^{\Phi(q) (Z(t)-Z(0)) -qt}.$$
On the new probability space $(\Omega, \mathcal{F}, \{\mathcal{F}_t\}_{t\geq 0}, \P^{\Phi(q)})$ we have
$Z(t)=(u_2-x)+c_2 t-S^{(q)}(t)$ for
$$S^{(q)}(t)=\sum_{i=1}^{N^{\Phi(q)}_t} U_i^{\Phi(q)},$$ and
where $N^{\Phi(q)}_t$ is a Poisson process with intensity $\lambda_{q}=\lambda \tilde{F}(\Phi(q))$ for $\tilde{F}(\theta)=\int_0^\infty e^{-\theta x}f(x)\,dx$
and $U^{\Phi(q)}$ has a density function:
$$f_{q}(x)=e^{-\Phi(q)x}f(x)/\tilde{F}(\Phi(q));$$
see Asmussen \cite[Th. 4.8, p. 38]{Asbookruinprob} and Rolski et al. \cite{Rolski} for details.
Denote $\tau_{i}=\inf\{t\geq 0: X_i(t)<0\}$ ($i=1,2$).
\begin{lem}\label{piece}
We have,
\begin{equation}\label{identityfirst}
\Exp\left[e^{-q  \tau_{U}} \textrm{   } 1_{( \tau_{U}<\tau_{L})}\right]=e^{-\Phi(q)x}\frac{V^{(q)}(u_2-x)}{V^{(q)}(u_2)},
\end{equation}
where for $y\geq u_2-u_1$ and $R=\frac{u_2-u_1}{c_1-c_2}$,
\begin{eqnarray}
\lefteqn{V^{(q)}(y)=\int_0^{y+c_2R}\P^{\Phi(q)}(\tau_1>R, X_1(R)\in dz|X_1(0)=y-(u_2-u_1))}\nonumber\\
&&\hspace{4cm}\P^{\Phi(q)}(\tau_{2}=\infty|X_2(0)=z)\, dz.\label{splot}
\end{eqnarray}
\end{lem}
\begin{proof}
Note that from the Optional Stopping Theorem:
$$\Exp\left[e^{-q  \tau_{U}} \textrm{   } 1_{( \tau_{U}<\tau_{L})}\right]=e^{-\Phi(q)x}\P^{\Phi(q)}( \tau_{U}<\tau_{L}),$$
where we use the fact that $Z(t\wedge \tau_U)\leq u_2$. Introduce $V^{(q)}(y)=\P^{\Phi(q)}( \tau_{L}=\infty|Z(0)=y)$.
Using the Strong Markov Property
\begin{eqnarray*}V^{(q)}(u_2-x)=\P^{\Phi(q)}( \tau_{U}<\tau_{L})V^{(q)}(u_2),
\end{eqnarray*}
since $\P^{\Phi(q)}(\tau_{U}> \tau_{L}, \tau_{L}=\infty)=0$.
This completes the proof of (\ref{identityfirst}).
The identity (\ref{splot}) is also a consequence of the Markov property applied to $Z$ at time $R$ being the zero of the line $z=(u_2-u_1) -(c_1-c_2)t$.
\exit
\end{proof}

The terms appearing in (\ref{splot}) could be identified in the following way. The ruin probability
$\P^{\Phi(q)}(\tau_{2}<\infty|X_2(0)=z)=1-\P^{\Phi(q)}(\tau_{2}=\infty|X_2(0)=z)$ was analyzed in many papers.
The reader is referred to the books by Gerber \cite{Gerberbook}, Grandell \cite{Grandell}, Asmussen \cite{Asbookruinprob} and Rolski et al. \cite{Rolski}.
Define $g_j(x)=c_jf_q(c_jx)$ for $j=1,2$ and for $x>0$,
\begin{equation}\label{ftj}f^j_{t}(x)=\frac{\P^{\Phi(q)}(\frac{1}{c_j}S^{(q)}(t)\in dx)}{dx}=e^{-\lambda_q t}\sum_{i=1}^\infty\frac{(\lambda_q t)^i}{i!} g_j^{*i}(x) \textrm{  for $j=1,2$}.
\end{equation}
From the Beekman-Pollaczeck-Khinchine
formula we have
$$\P^{\Phi(q)}(\tau_{2}<\infty|X_2(0)=z)=\left(1-\frac{\lambda_q}{c_2} \int_0^\infty v f_q(v)\, dz\right)\int_{z/c_2}^\infty f^2_{v-z/c_2}(v)e^{\lambda(z/c_2-v)}\, dv;$$
see e.g. Lef\'{e}vre and Loisel \cite[Cor. 3.8]{Loisel}.
For the exponential claim size with parameter $\alpha$, we have $\lambda_q=\lambda\alpha/(\alpha+\Phi(q))$ and
$f_q(x)=\alpha_q e^{-\alpha_q x}$ for $\alpha_q=\alpha+\Phi(q)$ and
$\Phi(q)=\frac{(\lambda+q-c_2\alpha)+\sqrt{(\lambda+q-c_2\alpha)^2+4c_2q\alpha}}{2c_2}$.
Moreover,
$$\P^{\Phi(q)}(\tau_{2}<\infty|X_2(0)=z)=\frac{\lambda_q}{c_2\alpha_q}\exp\left\{-(\alpha_q-\lambda_q/c_2)z\right\};$$
see e.g. Asmussen \cite[Cor. 3.2, p. 63]{Asbookruinprob} and Asmussen \cite[Th. 9.1, p. 108]{Asmbook}.

Similarly, from the ballot theorem (see Borovkov \cite{Borovkov}, Picard and Lef\'{e}vre \cite{picard} and Lef\'{e}vre and Loisel \cite[Lem. 3.3]{Loisel})
we have the following lemma.
\begin{lem}
For $\varphi(z)=v-\frac{z}{c_1}+R$ and $v=(y-(u_2-u_1))/c_1$,
\begin{eqnarray}
\lefteqn{
\frac{1}{dz}\P^{\Phi(q)}(\tau_1>R, X_1(R)\in dz|X_1(0)=c_1v)}\nonumber
\\&&=f^1_{R}(\varphi(z))-e^{-\lambda_q z/c_1}f^1_{R-z/c_1}(\varphi(z))\nonumber\\&&\quad -\int_{v}^{\varphi(z)}
\frac{z}{c_1(R+v-w)}f^1_{R+v-w}(\varphi(z)-w)f^1_{w-v}(w)\,dw.\label{ballot}
\end{eqnarray}
\end{lem}

To summarize, to find the cumulative dividend payments in this case we use identity (\ref{1.5b}) which
is based on finding $\Exp\left[e^{-q  \tau_{U}} \textrm{   } 1_{( \tau_{U}<\tau_{L})}\right]$ identified
in Lemma \ref{piece} via $\P^{\Phi(q)}(\tau_{2}=\infty|X_2(0)=z)$ and $\P^{\Phi(q)}(\tau_1>R, X_1(R)\in dz|X_1(0)=c_1v)$
for some $v$. Both these quantities could be found using function $f_t^j$ ($j=1,2$) given in (\ref{ftj}).
In contrast to the case $u_1>u_2$, a more explicit expression for the expected value of the dividend payments is impossible to derive.
Although De Vylder and Goovaerts \cite{Vylder1, Vylder2} show how to analyze numerically expressions of type (\ref{ballot}),
calculating expected dividend payments $V_1(u_1,u_2)$ for $u_1<u_2$ still leads to complications when identifying (\ref{pochodnatau}) and (\ref{splot}).

\sec{Conclusions}\label{concl}

In this paper we analyze joint dividend payments for two-dimensional risk process, when the reserves of two insurance companies are
related with each other through a proportional reinsurance. We consider two control mechanisms: refracting at linear barrier and impulse control.

In the first case we derive partial differential equations for the $n$th moment of the cumulative dividend payments.
We find also an explicit expression for the expected value of dividend payments when arriving claims have an exponential
distribution and the risk process is reflected at a linear barrier. It appeared that in contrast to the one-dimensional case this quantity
is given by a more complex expression, being a linear combination of two series. Besides it, the numerical analysis shows that the optimal choice of the barrier
(its upper left end $(0,b)$ and its slope $a$) depends
on the initial reserves of both companies. Thus the dependence of the two companies might affect the choice of the optimal strategy.

The impulse control with positive transaction costs produces more complex expressions for the mean of the dividend payments. The expression for the case when
the first company
has less initial reserves than the second company turned out to be much harder to analyze.
This is imposed on the way the dividends are paid: the payments (reducing reserves to the fixed level) are always made by the first company
and they are realized when the second company reaches fixed level of reserves. Thus, reserves of the second company
offer us a control mechanism in this case.

Extensions would be
with regard of  a penalty function taking
into account the severity of ruin, other ruin times or other types of the dependencies between reserves of insurance
companies. In particular, in our model claims for both companies arrive at the same time which "graphically"
induces jumps only in one direction.
The dependence structure could be relaxed by
allowing some claims to arrive to only one company.
Nevertheless, we leave this point for future research, since it seems that our methodology cannot be applied straightforwardly
to that extended framework. Even the simple case of a linear barrier strategy will result in more involved partial differential equations.

\section*{Acknowledgements}
This work is partially supported by the Ministry of Science and Higher Education of Poland under the grants N N201 394137 (2009-2011)
and N N201 525638 (2010-2011).

\end{document}